\documentclass[12pt]{article}
\usepackage{epsfig}
\usepackage{axodraw}
\def\theequation{\thesection.\arabic{equation}}
\def\KK{$\bar{K}^0$-$K^0$~}
\def\BB{$\bar{B}^0$-$B^0$~}
\def\BBs{$\bar{B}_s^0$-$B_s^0$~}

\def\BBds{$\bar{B}_{d,s}^0$-$B_{d,s}^0$~}

\def\simlt{\stackrel{<}{{}_\sim}}
\def\simgt{\stackrel{>}{{}_\sim}}

\newcommand{\bea}{\begin{eqnarray}}
\newcommand{\eea}{\end{eqnarray}}
\newcommand{\bd}{\begin{displaymath}}
\newcommand{\ed}{\end{displaymath}}
\newcommand{\be}{\begin{equation}}
\newcommand{\ee}{\end{equation}}

\def\nnb{\nonumber}
\def\bbuildrel#1_#2^#3{\mathrel{\mathop{\kern 0pt#1}\limits_{#2}^{#3}}}
\def\slash#1{\setbox0=\hbox{$#1$}#1\hskip-\wd0\dimen0=5pt\advance
       \dimen0 by-\ht0\advance\dimen0 by\dp0\lower0.5\dimen0\hbox
         to\wd0{\hss\sl/\/\hss}}

\newcommand{\TT}{\rule[-2mm]{0mm}{7mm}}

\def\theequation{\thesection.\arabic{equation}}

\newcommand{\vcb}{|V_{cb}|}

\newcommand{\vub}{|V_{ub}/V_{cb}|}

\renewcommand{\baselinestretch}{1.3}

\newcommand{\beq}{\begin{equation}}
\newcommand{\eeq}{\end{equation}}

\def\con{\ifmmode \hbox{\bf*} \else{\bf*}\fi}   
\def\scon{\ifmmode \hbox{\footnotesize\rm\bf*} \else{\footnotesize\rm\bf*}\fi}

\def\0#1{\relax\ifmmode\mathaccent"7017{#1}
        \else\accent23#1\relax\fi}              

\renewcommand{\baselinestretch}{1.2}

\begin{document}
\thispagestyle{empty}

{\normalsize\sf
\rightline {hep-ph/0107048}
\rightline{TUM-HEP-419/01}
\rightline{IFT-01/17}
\vskip 3mm
\rm\rightline{July 2001}
}

\vskip 5mm

\begin{center}
  
  {\LARGE\bf \boldmath{$\Delta M_s/\Delta M_d$},
    \boldmath{$\sin 2\beta$} and the angle \boldmath{$\gamma$} \\[2mm]
    in the Presence of New $\Delta F=2$ Operators}

\vskip 10mm

{\large\bf Andrzej J.~Buras$^1$, Piotr H.~Chankowski$^2$,}\\
{\large\bf Janusz Rosiek$^{1,2}$ and {\L}ucja S{\l}awianowska$^2$} \\[5mm]

{\small $^1$ Physik Department, Technische Universit{\"a}t M{\"u}nchen,}\\
{\small D-85748 Garching, Germany}\\
{\small $^2$ Institute of Theoretical Physics, Warsaw University}\\
{\small Ho\.za 69, 00-681 Warsaw, Poland}

\end{center}

\vskip 5mm

\renewcommand{\baselinestretch}{1.1} 
\begin{abstract}
\vskip 3mm {\small We present formulae for the mass differences $\Delta M_d$ 
and $\Delta M_s$ in the \BBds systems and for the CP violation parameter 
$\varepsilon$ which are valid in minimal flavour violation models giving rise 
to new four-fermion $\Delta F=2$ operators. Short distance contributions to 
$\Delta M_s$, $\Delta M_d$ and $\varepsilon$ are parameterized by three {\it
real} functions $F^s_{tt}$, $F^d_{tt}$ and $F^\varepsilon_{tt}$, respectively 
($F^s_{tt}=F^d_{tt}=F^\varepsilon_{tt}$ holds only if the Standard Model 
$(V-A)\otimes(V-A)$ operators dominate). We present simple strategies 
involving the ratio $\Delta M_s/\Delta M_d$, $\sin2\beta$ and $\gamma$ that 
allow to search for the effects of the new operators. We point out that their 
sizable contributions to the ratio $\Delta M_s/\Delta M_d$ would in principle 
allow $\gamma$ to be larger than $90^\circ$. Constraints on the functions 
$F^i_{tt}$ imposed by the present (and future) experimental data are also 
discussed. As an example we show that for large $\tan\bar\beta\equiv v_2/v_1$ 
and $H^+$ not too heavy, $F^s_{tt}$ in the MSSM with heavy sparticles can be 
substantially smaller than in the SM due the charged Higgs box contributions 
and in particular due to the growing like $\tan^4\bar\beta$ contribution of 
the double penguin diagrams involving neutral Higgs boson exchanges. As a 
result the bounds on the function $F^s_{tt}$ can be violated which allows 
to exclude large mixing of stops. In this scenario the range of $\sin2\beta$ 
following from $\varepsilon$ and $\Delta M_d$ is identical to the SM ones 
($0.5<\sin 2\beta<0.8$). On the other hand $\gamma$ following from 
$\Delta M_s/\Delta M_d$ is lower.}
\end{abstract}
\renewcommand{\baselinestretch}{1.2}

\newpage \setcounter{page}{1} \setcounter{footnote}{0}
\section{Introduction}
\setcounter{equation}{0} 

The determination of the CKM parameters and of the related unitarity
triangle (UT) are the hot topics in particle physics \cite{BUER}. In
this context a clean measurement of the angle $\beta$ in the unitarity
triangle through the time dependent CP asymmetry, $a_{\psi K_{\rm S}}(t)$, 
in $B^0_d(\bar B^0_d)\to \psi K_S$ decays is very important.

In the Standard Model (SM)
\be\label{a1}
a_{\psi K_{\rm S}}(t)\equiv -\,a_{\psi K_{\rm S}}\sin(\Delta M_d t)
=-\sin 2\beta \sin(\Delta M_d t),
\ee
thereby allowing direct extraction of $\sin 2\beta$.  The most recent
measurements of $a_{\psi K_{\rm S}}$ from the BaBar and Belle
Collaborations give
\begin{equation}\label{sinb}
a_{\psi K_{\rm S}} =\left\{ \begin{array}{ll}
0.59 \pm 0.14 \pm 0.05 & {\rm (BaBar)} ~\cite{BaBarNew} \\
0.99 \pm 0.14 \pm 0.06 & {\rm (Belle)} ~\cite{BelleNew}.
\end{array} \right.
\end{equation}
Combining these results with earlier measurements at CDF
$(0.79^{+0.41}_{-0.44})$ \cite{CDF} and by the ALEPH collaboration
$(0.84^{+0.82}_{-1.04}\pm 0.16)$ \cite{ALEPH} would give the grand average
\be
a_{\psi K_{\rm S}}=0.79\pm 0.10~,
\label{ga}
\ee
but in view of the fact that BaBar and Belle results are not fully consistent 
with each other we believe that a better description of the present situation 
is $a_{\psi K_{\rm S}}=0.80\pm 0.20$.

Similarly important for the determination of the unitarity triangle
will be the measurement of the ratio of the mass differences in the
\BBds systems, $\Delta M_s/\Delta M_d$.  The experimental values of
$\Delta M_d$ and $\Delta M_s$ read \cite{STO}
\be
\Delta M_d=(0.487\pm0.009) /ps~, \qquad
\Delta M_s\ge 15.0/ps~,
\ee
implying
\be
\frac{\Delta M_s}{\Delta M_d}\ge 30/ps~,
\ee
which is compatible with the SM expectations. Because theoretically
this ratio is considerably cleaner than $\Delta M_s$ and $\Delta M_d$
themselves, its precise measurement will have an important impact on
the determination of the unitarity triangle and on the tests of the SM
and its various extentions. As emphasized in
\cite{BUER,ALLO,BULAOS,BUGAGOJASI,BUBU} this impact will be even
stronger in conjunction with the measurement of $a_{\psi K_{\rm S}}$.
It is therefore exciting that $\Delta M_s/\Delta M_d$ should be
measured already this year in Run II at Fermilab, while improvements
of the $a_{\psi K_{\rm S}}$ measurements are expected from BaBar,
Belle, CDF, D0 and, at later stages, from BTeV and LHC experiments.

The result (\ref{ga}) for $a_{\psi K_{\rm S}}$ should be compared with
the value of $\sin 2\beta$ obtained from the analyses of the unitarity
triangle in the framework of the Standard Model (SM)
\cite{BUER,ALLO,SM} that center around $(\sin 2\beta)_{\rm SM}\approx
0.70$ with estimated errors ranging from $0.07$ to $0.24$. Clearly in
view of large experimental error in (\ref{ga}) and the considerable
uncertainty in the error estimates of $(\sin2\beta)_{\rm SM}$, the SM
fits are compatible with the experimental value of $a_{\psi K_{\rm S}}$. 
The small value of $a_{\psi K_{\rm S}}$ found earlier by the BaBar 
collaboration \cite{BaBar} has not been confirmed by most recent data.
On the other hand, the large value of $a_{\psi K_{\rm S}}$ measured by
the Belle collaboration may be suggestive of new physics contributions 
to $B^0_d(\bar B^0_d)\to\psi K_S$, \BBds mixing, \KK mixing and/or the
parameter $\varepsilon$ measuring CP violation in the \KK mixing. Thus 
the analyses \cite{BUBU}, \cite{EYNIPE}--\cite{ALLU} of new contributions 
done in the context of small $a_{\psi K_{\rm S}}$ value reported by BaBar 
\cite{BaBar} could still be relevant if properly reformulated. Such new 
contributions could modify not only the relation between 
$a_{\psi K_{\rm S}}$ and $\sin 2\beta$ in (\ref{a1}) but also the value 
of $\sin 2\beta$ obtained from the fits of the unitarity triangle.

In general models of new physics potentially contributing to $a_{\psi
K_{\rm S}}$, $\Delta M_{d,s}$ and/or $\varepsilon$ fall into the two
following broad classes \cite{BUER}:
\begin{itemize}
\item Models in which the CKM matrix remains the unique source of
  both, flavour and CP violation. The effects of this source are
  however modified by the new interaction vertices (of new particles)
  in which the CKM matrix elements appear.
\item Models with entirely new sources of flavour and/or CP violation.
\end{itemize}
The first class can be conveniently further subdivided into the
so-called MFV models \cite{CIDEGAGI,BUGAGOJASI} and the generalized
MFV models (GMFV).

The characteristic feature of the MVF models (MVF scenarios of new
physics) is the strong dominance in their low energy effective
Hamiltonian of the same operators that occur in the low energy
effective Hamiltonian of the SM. In such models the formula (\ref{a1})
remains valid and the relation between the ratio $\Delta M_s/\Delta
M_d$ and the length of one side of the unitarity triangle, $R_t$, is
as in the SM: it remains independent of the parameters of the
particular model. Thus, for this class of models the unitarity
triangle is universal \cite{BUGAGOJASI}. The distinction between
different models of this class can then be made through the study of
$\varepsilon$ and $\Delta M_d$ which in contrast to 
$a_{\psi K_{\rm S}}$ and $\Delta M_s/\Delta M_d$ do depend explicitly 
on new physics contributions.  A detailed analysis of the profile of 
the UT in supersymmetric scenarios of this class can be found in 
\cite{ALLO}. Other discussions of the MFV models can be found in
\cite{MAVI,BEPE,BUFL}.

The GMFV models generalize the MFV models by allowing for significant
contributions of the nonstandard operators in the effective low energy
Hamiltonian. In this class of models the formula (\ref{a1}) is still
valid but the relation between $\Delta M_s/\Delta M_d$ and $R_t$ is
modified.  Hence, $R_t$ determined from the measured ratio $\Delta
M_s/\Delta M_d$ does depend on the parameters of the model.

While the presence of new CP violating phases in \BBds mixing and \KK
mixing could turn out to be necessary to explain the future precise value
of $a_{\psi K_{\rm S}}$, it is important to investigate first the scenarios 
that do not invoke new sources of flavour and/or CP violation. It is 
therefore useful to analyze first MFV and GMVF models that are more 
constrained than the more general scenario mentioned above.

It turns out that in the MFV models there exists an {\it absolute}
lower bound on $\sin 2\beta$ \cite{BUBU} that follows from the
interplay of $\Delta M_d$ and $\varepsilon$ and depends mainly on
$\vcb$, $\vub$ and the non-perturbative parameters $\hat B_K$,
$F_{B_d}\sqrt{\hat B_{B_d}}$ entering the analysis of the unitarity
triangle. An updated conservative lower bound on $\sin2\beta$ obtained
by scanning independently all relevant input parameters reads \cite{BUER}
\be
\label{bound}
(\sin 2\beta)_{\rm min}=0.42~.
\ee
As analyzed in \cite{BUBU,BUER}, this bound could be considerably
improved when the values of $\vcb$, $\vub$, $\hat B_K$,
$F_{B_d}\sqrt{\hat B_{B_d}}$, will become better known or if $\Delta M_s$
is measured so that the ratio $\Delta M_s/\Delta M_d$ can be used,
along with the non-perturbative parameter $\xi$, to determine the
length of one side of the UT. The lower bound (\ref{bound}) is fully
consistent with the experimental data but as the latter are not yet very
precise it could prove useful when the knowledge of $|V_{cb}|$,
$|V_{ub}/V_{cb}|$, $\hat B_K$, $F_{B_d}\sqrt{\hat B_{B_d}}$, $\Delta M_s$
and $\xi$ improves. Note that the bound (\ref{bound}) allows values of
$\sin2\beta$ that are slightly smaller than the ones obtained from the
fits to the unitarity triangle within the SM, $(\sin2\beta)_{\rm SM}>0.5$.
In view of the unexpectedly high value of $a_{\psi K_{\rm S}}$ found by
the Belle collaboration \cite{BelleNew}, more interesting at present appears
the upper bound on $\sin2\beta$ in MFV models that reads \cite{BULAOS,BUFL}
\be
(\sin2\beta)_{\rm max}=2R_b\sqrt{1-R_b^2}~.
\ee
Here $R_b$ is the lenght of one side of the unitarity triangle (see 
fig.~\ref{fig:bcrs1}) given in terms of $|V_{ub}/V_{cb}|$ in 
eq.~(\ref{eqn:Rb}). With the input parameters specified in table 1, one 
obtains $(\sin2\beta)_{\rm max}=0.82$ that is fully consistent with the 
BaBar result \cite{BaBarNew} but appears to be violated by the Belle result
\cite{BelleNew}. We will return to this issue in the course of this paper.
The natural next step is to
exploit GMFV models.  In the present paper we would like to make this
step and present general formulae relevant for the analysis of the
unitarity triangle and $\sin 2\beta$ in the GMVF models.

Examples of the MFV models are the Two Higgs Doublet Models II
(2HDM(II)) and the MSSM, in which sfermion mass squared matrices are
aligned with the corresponding fermion mass matrices and the CP
violating phases of the gaugino masses, $\mu$ and $A$ parameters are
all set to zero, provided the ratio of the vacuum expectation values
of the two Higgs doublets, $v_2/v_1\equiv\tan\bar\beta$ is not too
large. It is well known \cite{MIPORO,BUGAGOJASI2} that in both these
models the contribution of light charged Higgs boson and/or (in the
case of supersymmetry) charginos and stops to the Wilson coefficient
of the standard $(V-A)\otimes(V-A)$ operator can significantly enhance
the $tW^\pm$ contribution to $\Delta M_s$, $\Delta M_d$ and
$\varepsilon$. In this paper we show that for large
$\tan\bar\beta\equiv v_2/v_1$ both models become GMFV models. In
particular in the MSSM in the limit $M_W\simlt M_{H^+}\ll M_{\rm
  sparticle}$, which we consider here for simplicity, we find that:
\begin{itemize}
\item There can be significant contributions to $\Delta M_s$ from the
  charged Higgs box diagrams (growing like $\tan^2\bar\beta$ at the
  1-loop level and faster after including leading higher order
  corrections) and, growing as $\tan^4\bar\beta$, contributions
  arising from the double penguin diagrams involving the neutral Higgs
  scalars.  Compared to the contribution of the extended Higgs and
  chargino/stop sectors relevant for low $\tan\bar\beta$, the
  interesting feature of all these new contributions is their sign
  which is opposite to the standard $tW^\pm$ box contribution.
\item Compared to $\Delta M_s$, the corresponding contributions to
  $\Delta M_d$ and $\varepsilon$ are suppressed by the quark mass
  ratios $m_d/m_s$ and $m_d/m_b$, respectively.
\item Consequently, in this scenario $\sin2\beta$ cannot deviate
  significantly from its SM value, i.e. the lower bound (\ref{bound})
  can never be reached.
\item Present experimental data strongly limit large mixing of stops
  if their mass difference is large compared to the electroweak scale.
\item If $a_{\psi K_{\rm S}}$ is found below $0.5$ or above $0.82$ this 
  particular
  supersymmetric scenario will be disfavoured (together with the SM).
  If $a_{\psi K_{\rm S}}\approx0.7$ this scenario can lead to the
  $\gamma$ angle slightly smaller (depending on the measured value of
  $\Delta M_s$) than in the SM.
\end{itemize}
The full MSSM with large $\tan\bar\beta$, including the effects of
light sparticles, will be analyzed in the forthcoming paper
\cite{BUCHROSL}.

Our paper is organized as follows. In section 2 we generalize the MFV
formulae of \cite{BUBU,BUGAGOJASI,BUER} to GMFV models. While in MFV
models the new physics short distance contributions to \BBds mixing
and $\varepsilon$ can be described by only a single function $F_{tt}$,
the transition to the GMFV models (in which new operators contribute)
requires the introduction of three {\it real} functions $F^d_{tt}$,
$F^s_{tt}$ and $F^\varepsilon_{tt}$.  We present simple strategies
involving the ratio $\Delta M_s/\Delta M_d$, $\sin 2\beta$ and the
angle $\gamma$ that allow to search for the effects caused by new
operators (sec. 2.3) and discuss model independent bounds on the
functions $F^{s,d,\varepsilon}_{tt}$ which follow from the present and
future experimental data (sec. 2.4). We also point out that in this
class of models the function $F^s_{tt}$ can be directly measured
through $\Delta M_s$ and that the angle $\gamma$ could be larger than
$90^\circ$. The formulae necessary to express the functions
$F^{s,d,\varepsilon}_{tt}$ directly in terms of the Wilson
coefficients of the four-fermion operators are collected in Section 3.
The 2HDM(II) and the MSSM with large $\tan\bar\beta\equiv v_2/v_1$ and
heavy sparticles are discussed in Section 4.  We give complete
formulae for the one loop contribution of the box diagrams involving
charged Higgs bosons and derive simple approximate formulae describing
the dominant effects of the double penguin diagrams.  Consequences of
their large contribution and implications of the bounds presented in
sec. 2.4 are also discussed. We conclude in Section 5.

\section{Basic Formulae}
\setcounter{equation}{0}
\subsection{Effective Hamiltonian in GMVF Models}
The effective weak Hamiltonian for $\Delta F=2$ transitions in the
GMVF models can be written as follows
\be\label{heff}
H_{\rm eff}^{\Delta {\rm F}=2} = {G_F^2M_W^2\over16\pi^2}
\sum_i V^i_{\rm CKM} C_i(\mu) Q_i~.
\ee
Here $Q_i$ are $\Delta F=2$ operators, $G_F$ is the Fermi constant and
$V^i_{\rm CKM}$ is the appropriate {Cabibbo-Kobayashi-Maskawa} (CKM)
factor.  Because in the GMFV models the CKM matrix is by definition
the only source of flavour and CP violation, the Wilson coefficients
$C_i(\mu)$ are {\it real}. Using this Hamiltonian with the Wilson
coefficients evaluated at the appropriate scale $\mu$ one can
calculate $\Delta F=2$ amplitudes, in particular the mass differences
$\Delta M_{d,s}$ and the CP violation parameter $\varepsilon$ measured
in $K\to\pi\pi$ decays.

The full set of dimension six operators contributing to $\Delta F=2$ 
transitions consists of 8 operators. According to the chirality of the 
quark fields they can be split into 5 separate sectors.
The operators belonging to the VLL, LR and SLL sectors read:
\bea
Q_1^{\rm VLL} &=& (\bar{d}_J\gamma_\mu P_L d_I)
                  (\bar{d}_J\gamma^\mu P_L d_I),
\nnb\\[4mm]
Q_1^{\rm LR} &=&  (\bar{d}_J\gamma_\mu P_L d_I)
                  (\bar{d}_J\gamma^\mu P_R d_I),
\nnb\\
Q_2^{\rm LR} &=&  (\bar{d}_J P_L d_I)
                  (\bar{d}_J P_R d_I),
\nnb\\[4mm]
Q_1^{\rm SLL} &=& (\bar{d}_J P_L d_I)
                  (\bar{d}_J P_L d_I),
\nnb\\
Q_2^{\rm SLL} &=& (\bar{d}_J\sigma_{\mu\nu}P_L d_I)
                  (\bar{d}_J\sigma^{\mu\nu}P_L d_I),
\label{eqn:ops}
\eea
where $I,\,J$ are the flavour indices (i.e. $d_3\equiv b$, $d_2\equiv
s$, $d_1\equiv d$ and, analogously, $u_3\equiv t$, $u_2\equiv c$,
$u_1\equiv u$), $\sigma_{\mu\nu}={1\over2} [\gamma_\mu,\gamma_\nu]$,
$P_{L,R} ={1\over2}(1\mp\gamma_5)$ and the colour indices are
contracted within the brackets. The operators belonging to the VRR and
SRR sectors are obtained from $Q_1^{\rm VLL}$ and $Q_i^{\rm SLL}$ by
interchanging $P_L$ and $P_R$. Since QCD preserves chirality, there is
no mixing between different sectors. Moreover, the QCD evolution
factors from high energy to low energy scales in the VRR and SRR
sectors are the same as in the VLL and SLL sectors, respectively.
However, one should remember that the initial conditions $C_i(\mu_t)$
(where $\mu_t={\cal O}(m_t)$) are in general different for operators
involving $P_L$ and $P_R$. In the limit in which the effective
Hamiltonian (\ref{heff}) is dominated by the single $Q_1^{\rm VLL}$
operator one recovers the results of the MFV models.

The QCD renormalization group factors relevant for the Hamiltonian
(\ref{heff}) have been calculated at the NLO level in
\cite{BUJAWE,URKRJESO,CET0,BUMIUR,CET,BUJAUR} where the last four
papers deal with the LR and SLL(SRR) operators. In particular in ref.
\cite{BUJAUR} master formulae for $\Delta F=2$ NLO QCD factors
relating $C_i(\mu_t)$ to $C_i(\mu)$ where $\mu_t={\cal O}(m_t)$ and
$\mu={\cal O}(m_b)$ or $\mu={\cal O}(2$~GeV) have been presented and
evaluated numerically in the NDR renormalization scheme. Below we will
exploit the general formulae of ref. \cite{BUJAUR} expressing $\Delta
M_d$, $\Delta M_s$ and $\varepsilon$ in terms of the non-perturbative
parameters $B_i$.

\begin{figure}[htbp]
\begin{center}
\epsfig{file=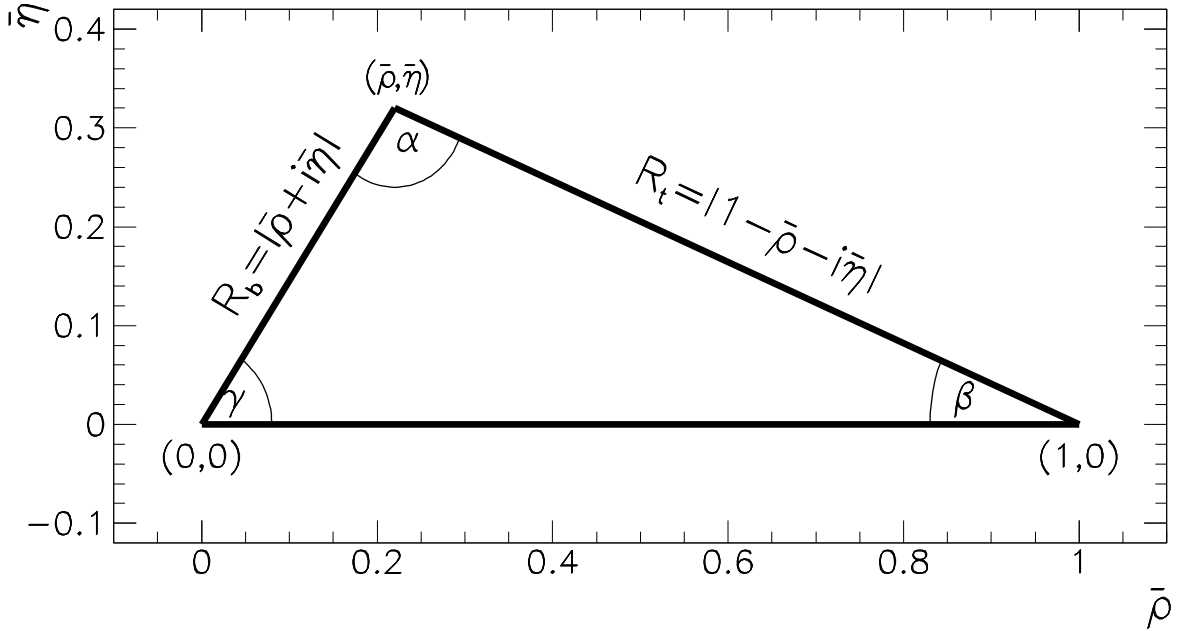,width=\linewidth}
\end{center}
\caption{Unitarity Triangle.}
\label{fig:bcrs1}
\end{figure}

\subsection{\boldmath{$\Delta M_d$}, \boldmath{$\Delta M_s$}, 
\boldmath{$\varepsilon$} and \boldmath{$\sin2\beta$} in the GMVF Models}

It is straightforward to generalize the formulae of refs.
\cite{BUER,BUBU,BUGAGOJASI,BUFL} to the case of GMFV models.  To this
end, following the notation of ref. \cite{ALLO}, we introduce three
real functions
\be
F^d_{tt}=S_0(x_t) \lbrack 1+f_d\rbrack~,
\qquad
F^s_{tt}=S_0(x_t) \lbrack 1+f_s\rbrack~,
\qquad
F^\varepsilon_{tt}=S_0(x_t) \lbrack 1+f_\varepsilon\rbrack~,
\ee
relevant for $\Delta M_d$, $\Delta M_s$ and $\varepsilon$
respectively.  $S_0(x_t)$ with $x_t=m_t^2/M_W^2$ is the function
resulting from box diagrams with $(t,W^\pm)$ exchanges.
$S_0(x_t)\approx2.38\pm0.11$ for $\bar m_t(m_t)=(166\pm5)$ GeV. In
order not to complicate the expressions below we assume $F_{tt}^i>0$.
Generalization to negative $F_{tt}^i$ can be easily done following the
discussion of ref. \cite{BUFL}.  We further split the parameters $f_i$
into universal and non-universal parts

\be\label{uni}
f_i=f_{\rm uni}+\tilde f_i
\ee
where $\tilde f_i=0$ in the MFV models \cite{BUBU}. We have then,
\begin{equation}
\Delta M_q = {G_{\rm F}^2M_W^2\over6\pi^2}M_{B_q} \eta_B 
\hat B_{B_q} F_{B_q}^2  |V_{tq}|^2 F_{tt}^q,
\qquad q=d,s
\label{eqn:xds}
\end{equation}
where $F_{B_q}$ is the $B_q$-meson decay constant, $\hat B_{B_q}$ is a
non-perturbative parameter and $\eta_B=0.55$ is the QCD factor
\cite{BUJAWE,URKRJESO}. The measurements of $\Delta M_q$ determine the
length $R_t$ of one side of the unitarity triangle (shown in fig.
\ref{fig:bcrs1}) defined by
\begin{equation}\label{eqn:rtdef}
R_t\equiv\frac{|V_{td}^{}V^*_{tb}|}{|V_{cd}^{}V^*_{cb}|} =
\sqrt{(1-\bar\varrho)^2 +\bar\eta^2}
={1\over\lambda}\left|\frac{V_{td}}{V_{cb}}\right|~.
\end{equation}
Here $\bar\varrho=\varrho (1-\lambda^2/2)$,
$\bar\eta=\eta(1-\lambda^2/2)$ \cite{BULAOS}, and $\lambda$, $\varrho$
and $\eta$ are the Wolfenstein parameters \cite{WO}. As in ref.
\cite{BUFL} we set $\lambda=0.222$ in the analytic formulae
below.\footnote{Because of that some numerical factors in the formulae
  below differ from their counterparts in refs. \cite{BUER,BUBU} where
  $\lambda=0.220$ has been used.  This change has only a very small
  impact on the numerical analysis.  In particular the bound
  (\ref{bound}) remains unchanged. On the other hand the increased
  value of $\lambda$ shifts $V_{ud}$ closer to its experimental value.
  See ref. \cite{SM} for the discussion of this point.}  Other input
parameters are collected in table~\ref{tab:inputparams}.

\begin{table}[thb]
\caption[]{The ranges of the input parameters.
\label{tab:inputparams}}
\vspace{0.4cm}
\begin{center}
\begin{tabular}{|c|c|c|}
\hline
{\bf Quantity} & {\bf Central value} & {\bf Error}\\
\hline
$\lambda$ & 0.222 & $\pm 0.0018 $ \\
$|V_{cb}|$ & 0.041 & $\pm 0.002$ \\
$\vub$ & $0.085$ & $\pm 0.018 $ \\
$|V_{ub}|$ & $0.00349$ & $\pm 0.00076$ \\
$\hat B_K$ & 0.85 & $\pm 0.15$ \\
$\varepsilon$ & $2.280\times10^{-3}$ & $\pm0.013\times10^{-3}$ \\
$\sqrt{\hat B_{B_d}} F_{B_d}$ & $230 ~{\rm MeV}$ & $\pm 40~{\rm MeV}$ \\
$\sqrt{\hat B_{B_s}} F_{B_s}$ & $265 ~{\rm MeV}$ & $\pm 40~{\rm MeV}$ \\
$\xi$ & $1.15$ & $\pm 0.06$ \\
$m_t$ & $166~{\rm GeV}$ & $\pm 5~{\rm GeV}$ \\
$\Delta M_d$ & $0.487/\mbox{ps}$ & $\pm 0.009/\mbox{ps}$ \\
$\Delta M_s$  & $>15.0/\mbox{ps}$ & \\
$M_W$ & $80.4$ GeV & \\
\hline
\end{tabular}
\end{center}
\end{table}

{}From $\Delta M_d$ and $\Delta M_d/\Delta M_s$ we find
\begin{equation}\label{RT}
R_t= 1.084~\frac{R_0}{A}{1\over\sqrt{F^d_{tt}}},
\qquad
R_0\equiv\sqrt{\frac{\Delta M_d}{0.487/{\rm ps}}}
         \left[\frac{230~{\rm MeV}}{\sqrt{\hat B_{B_d}} F_{B_d}}\right]
         \sqrt{\frac{0.55}{\eta_B}}
\ee
and
\be\label{Rt}
R_t=0.819~\xi\sqrt{{\Delta M_d\over0.487/{\rm ps}}}
             \sqrt{{15/{\rm ps}\over\Delta M_s}}
             \sqrt{R_{sd}}~,
\ee
respectively where the Wolfenstein parameter A is defined by
$\vcb=A\lambda^2$. Here
\be\label{rsd}
R_{sd}=\frac{1+f_s}{1+f_d}, \qquad
\xi=\frac{F_{B_s} \sqrt{\hat B_{B_s}}}{F_{B_d} \sqrt{\hat B_{B_d}}}~.
\ee
The measurement of the parameter $\varepsilon$ imposes the constraint
which reads:
\begin{equation}\label{eqn:epsk}
\bar\eta \left[(1-\bar\varrho) A^2 \eta_2 F^\varepsilon_{tt}
+P_c(\varepsilon) \right] A^2 \hat B_K = 0.204~,
\end{equation}
where $\eta_2=0.57$ is the QCD factor \cite{BUJAWE} and
$P_c(\varepsilon)$ summarizes the contributions not proportional to
$V_{ts}^*V_{td}$.  With (\ref{RT}) and (\ref{eqn:epsk}), the formula
of ref.~\cite{BUBU} for $\sin 2\beta$ valid in MFV models generalizes to
\be\label{main}
\sin 2\beta=\frac{1.65}{ R^2_0\eta_2} R_{d\varepsilon}
\left[\frac{0.204}{A^2 B_K}
-\bar\eta P_c(\varepsilon)\right], \qquad
R_{d\varepsilon}=\frac{1+f_d}{1+f_\varepsilon}~.
\ee
Note that new physics can affect $\sin 2\beta$ both through $f_d$ and
$f_\varepsilon$ in (\ref{main}) and indirectly through $\bar\eta$. We
assume as in \cite{BUGAGOJASI,BUBU} that new physics contributions to
$P_c(\varepsilon)$ are negligible. In this case
$P_c(\varepsilon)=0.30\pm 0.05$ \cite{HENI}.

\subsection{How to distinguish GMFV from MFV}

In general, the presence of new $\Delta F=2$ operators causes the
departure from the relation $F^d_{tt}=F^s_{tt}=F^\varepsilon_{tt}$
valid in the MFV models. This means for instance that the ratio
$\Delta M_s/\Delta M_d$, being now dependent on new physics
contributions, cannot be used any longer for the construction of the
universal unitarity triangle \cite{BUGAGOJASI}. In other words the
dictionary between $R_t$ and $\Delta M_s/\Delta M_d$, as given by
(\ref{Rt}), differs from the corresponding one in the MFV models
because $R_{sd}\not=1$.  This fact offers a possibility to distinguish
experimentally between these two classes of models.  Two strategies
are presented below. Because of the unitarity of the CKM matrix these
strategies are related to each other.

\begin{figure}[htbp]
\begin{center}
\epsfig{file=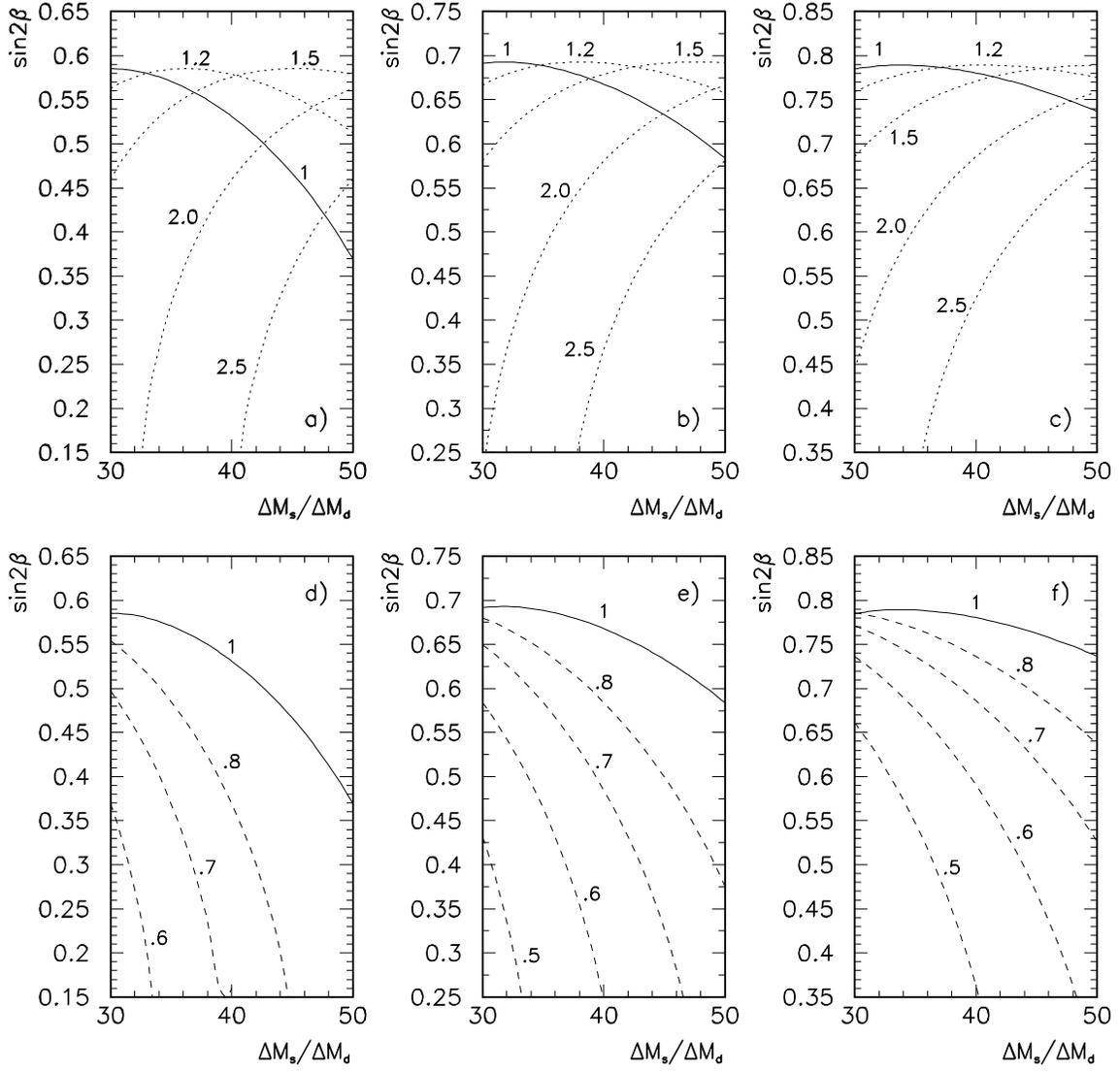,width=\linewidth}
\end{center}
\caption{\protect $\sin2\beta$ as a function of $\Delta M_s/\Delta M_d$ for
$\vub=0.070$ (panels a and d), ~0.085 (panels b and e), and 0.10 
(panels c and f) for different values of $R_{sd}$ 
(marked on the curves) and $\xi=1.15$.}
\label{fig:bcrs2}
\end{figure}

\subsubsection{Strategy A}
For given values of $\vub$ and $\Delta M_s/\Delta M_d$ one can determine
(see fig. \ref{fig:bcrs1})
\begin{equation}\label{eqn:Rb}
R_b \equiv \frac{| V_{ud}^{}V^*_{ub}|}{| V_{cd}^{}V^*_{cb}|}
= \sqrt{\bar\varrho^2 +\bar\eta^2}
= (1-\frac{\lambda^2}{2})\frac{1}{\lambda}
\left| \frac{V_{ub}}{V_{cb}} \right|,
\end{equation}
and $R_t$ by means of (\ref{Rt}), respectively. This
gives the apex of the unitarity triangle with
\begin{equation}\label{16a}
\bar\varrho = {1\over2}(1+R^2_b-R^2_t)~,
\quad\quad
\bar\eta=\sqrt{R_b^2-\bar\varrho^2}.
\end{equation}
and consequently 
\be\label{ss}
\sin 2\beta=\frac{2\bar\eta(1-\bar\varrho)}{R^2_t}~.
\ee
These formulae establish the relation between $\sin2\beta$ and 
$\Delta M_s/\Delta M_d$ that depends on the ratio $R_{sd}$. In 
fig.~\ref{fig:bcrs2} we show $\sin2\beta$  as a function of 
$\Delta M_s/\Delta M_d$ for $\vub=0.070,~0.085,~0.10$ and different
values of the ratio $R_{sd}$. To this end we have set $\xi=1.15$. We observe
that for $\Delta M_s/\Delta M_d$ below $40$ the distinction between GMFV and
MFV models will be difficult unless the values of $\xi$, $\vub$,
$a_{\psi K_{\rm S}}$ and $\Delta M_s$ will be known very accurately or the
value of $R_{sd}$ differs substantially from unity. For larger values of
$\Delta M_s/\Delta M_d$ the distinction is clearer.

{}From fig.~\ref{fig:bcrs2} the impact of the measurement of $\Delta
M_s/\Delta M_d$ on the allowed values of $\sin2\beta$ is clearly seen.
For $0.8<R_{sd}<1.2 $ and $\Delta M_s/\Delta M_d<40$ only values 
compatible with current experimental result (\ref{ga}) and well
above the bound (\ref{bound}) are allowed.
On the other hand, for
sufficiently low or sufficiently large values of $R_{sd}$, smaller values 
of $\sin2\beta$ are also possible. While such low values of $\sin2\beta$ 
are still compatible with the BaBar result, they seem to be excluded by the 
Belle measurement of $a_{\psi K_{\rm S}}$. In fact, as follows from
fig.~\ref{fig:bcrs2}, the latter measurement can be accomodated only if
$R_{sd}\approx1$ and $|V_{ub}/V_{cb}|\approx0.1$.

It is also easy to find that the two possibilities, small $R_{sd}$ and
large $R_{sd}$, favour small and large values of the angle $\gamma$ in
the unitarity triangle, respectively.  Which of these two
possibilities is favoured by the data can only be decided by other
measurements. This includes $\varepsilon$, $\Delta M_d$ alone and in
particular a direct measurement of $\gamma$. This brings us to the
second strategy.

\subsubsection{Strategy B}
The angle $\gamma$ in the unitarity triangle can be found from $R_t$
and $\beta$ determined through (\ref{a1}) by using the relation
\be\label{gar}
\cot\gamma=\frac{1-R_t\cos\beta}{R_t\sin\beta}.
\ee
Expressing $R_t$ in terms of $\Delta M_s/\Delta M_d$ by means of
(\ref{Rt}) allows to calculate $\gamma$ as a function of $\sin
2\beta$, $\Delta M_s/\Delta M_d$ and $R_{sd}$. For $R_{sd}\not= 1$ the
predictions for $\gamma$ in GMFV models will generally differ from
those in the MFV models. Comparing these predictions with future
direct measurements of $\gamma$ it will be possible to distinguish
between these two classes of models and check whether the inclusion of
new operators is required by the data. In fig.~\ref{fig:bcrs3} we show
$\gamma$ as a function of $\Delta M_s/\Delta M_d$ for $\sin2\beta=0.4$,~
$0.6$, ~$0.8$ and various values of $R_{sd}$.

We observe that the distinction between MFV and GMFV models in this
strategy is, in contrast to strategy A, very transparent in the full
range of $\Delta M_s/\Delta M_d$ considered. As this strategy involves
only $a_{\psi K_S}$ and $\Delta M_s/\Delta M_d$, that are
theoretically cleaner than $\vub$, it is this strategy which in the
future should play the crucial role in the distinction between the MFV
and GMFV models. The ratio $\Delta M_s/\Delta M_d$ and the asymmetry
$a_{\psi K_S}$ should be determined very precisely in the coming
years. The determination of $\gamma$ is more difficult but should be
achieved at LHCb and BTeV.  Some information on the angle $\gamma$
should also be gained from the $B_d\rightarrow\pi K$ decays measured
by CLEO, BaBar and Belle and by the combination of
$B_d\rightarrow\pi^+\pi^-$ rate (already measured by these three
collaborations) and the rate of the $B_s\rightarrow K^+K^-$ decay
\cite{FL} which are going to be measured at Tevatron.

As seen in fig. \ref{fig:bcrs3} the values of $\gamma$ for $R_{sd}\le
1.2$ are below $90^\circ$. On the other hand for substantially higher
$R_{sd}$ also $\gamma >90^\circ$ is possible. The possibility of
$\gamma >90^\circ$ resulting from the unitarity triangle fits is very
interesting in view of several analyses
\cite{BUFL2,HEHOUYA,BEBENESA,DUYAZH} of two-body non-leptonic decays
$B\to\pi K,~\pi\pi$ that favour $\gamma >90^\circ$ in contradiction
with the usual unitarity triangle analyses that confidently give
$\gamma < 90^\circ$. With increasing $\Delta
M_s/\Delta M_d$ this problem will become more serious.

\begin{figure}[htbp]
\begin{center}
\epsfig{file=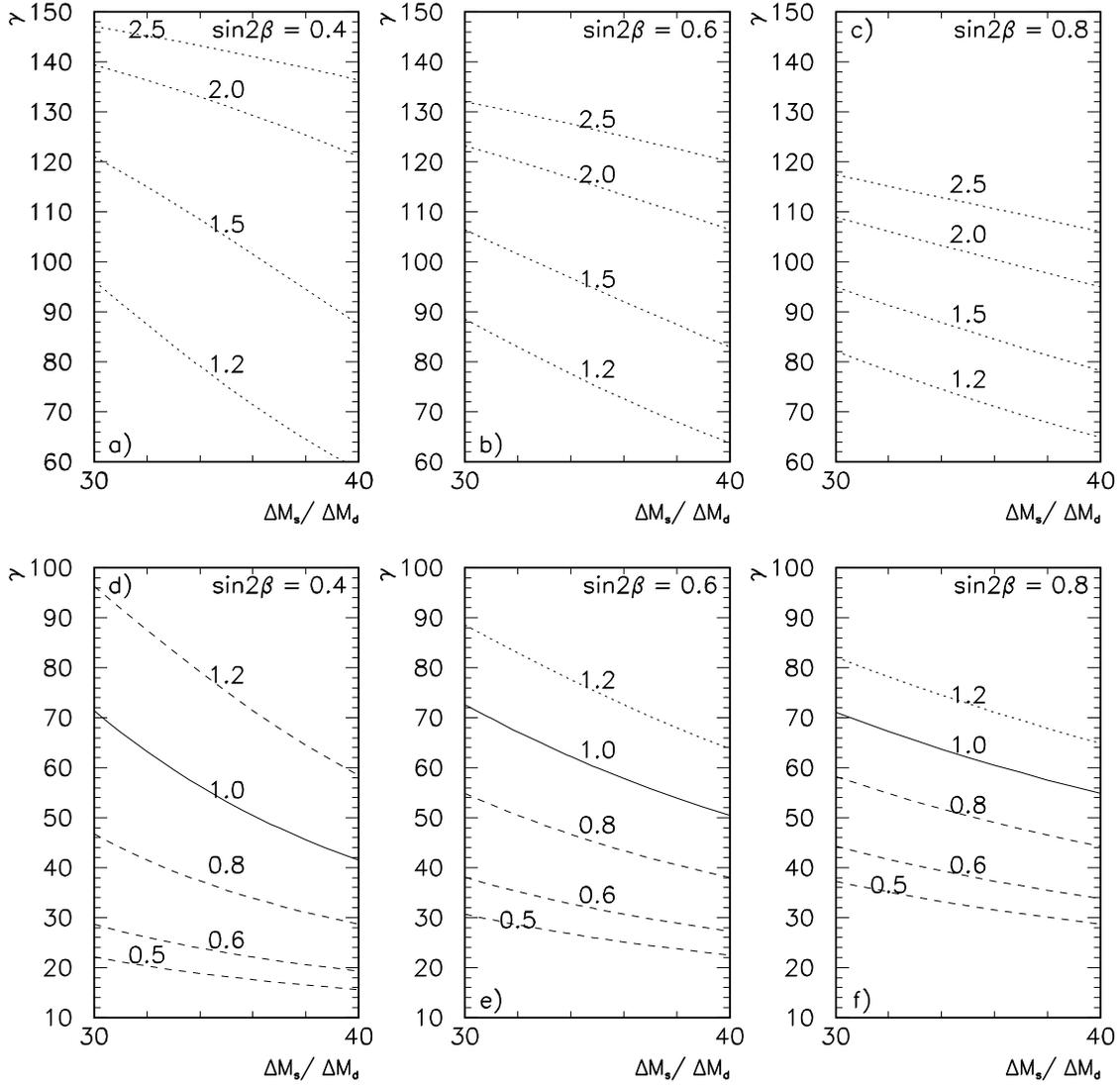,width=\linewidth}
\end{center}
\caption{$\gamma$ as a function of $\Delta M_s/\Delta M_d$ for 
$\sin 2\beta=0.4$, ~$0.6$, ~$0.8$ and various values of $R_{sd}$
(marked on the curves).}
\label{fig:bcrs3}
\end{figure}

In view of sizable theoretical uncertainties in the analyses of
$B\to\pi K,~\pi\pi$ and of large experimental errors in the
corresponding branching ratios it is not yet clear whether the
discrepancy in question is serious. For instance \cite{CIFRMAPISI}
sizable contributions of the so-called charming penguins to the
$B\to\pi K$ amplitudes could shift $\gamma$ extracted from these
decays below $90^\circ$ but at present these contributions cannot be
calculated reliably. Similar role could be played by annihilation
contributions and large non-factorizable $SU(3)$ breaking effects
\cite{BUFL2}.  Also, a new physics contribution in the electroweak
penguin sector could also shift $\gamma$ to the first quadrant
\cite{BUFL2}.  It should be however emphasized that the problem with
the angle $\gamma$, if it persisted, would put into difficulties not
only the SM but also the full class of MFV models in which the lower
bound on $\Delta M_s/\Delta M_d$ implies $\gamma < 90^\circ$. On the
other hand as seen in fig.~\ref{fig:bcrs3} for sufficiently high
values of $R_{sd}$, the angle $\gamma$ resulting from the unitarity
triangle analysis can easily be in the second quadrant provided
$\Delta M_s/\Delta M_d$ is not too large.

Clearly a general analysis of the unitarity triangle involving
$\varepsilon$, $\Delta M_{s,d}$, $\vub$ and $\vcb$ can also be used to
search for the effects caused by the new operators but the two
strategies outlined above have in our opinion the best chance to
distinguish between MFV and GMFV models in a transparent manner.

\begin{figure}[htbp]
\begin{center}
\begin{tabular}{p{0.72\linewidth}}
\mbox{\epsfig{file=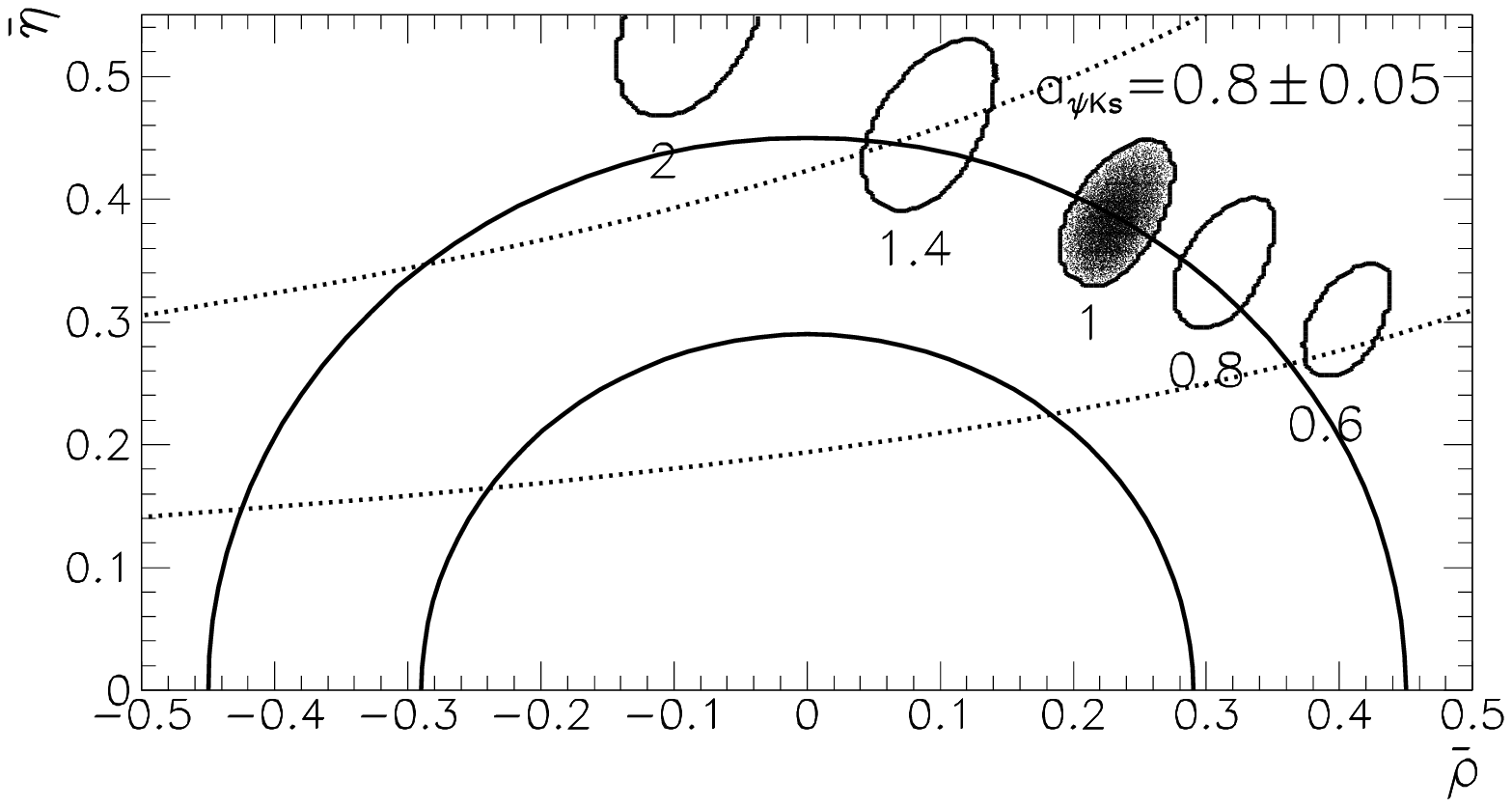,width=\linewidth}}\\
\mbox{\epsfig{file=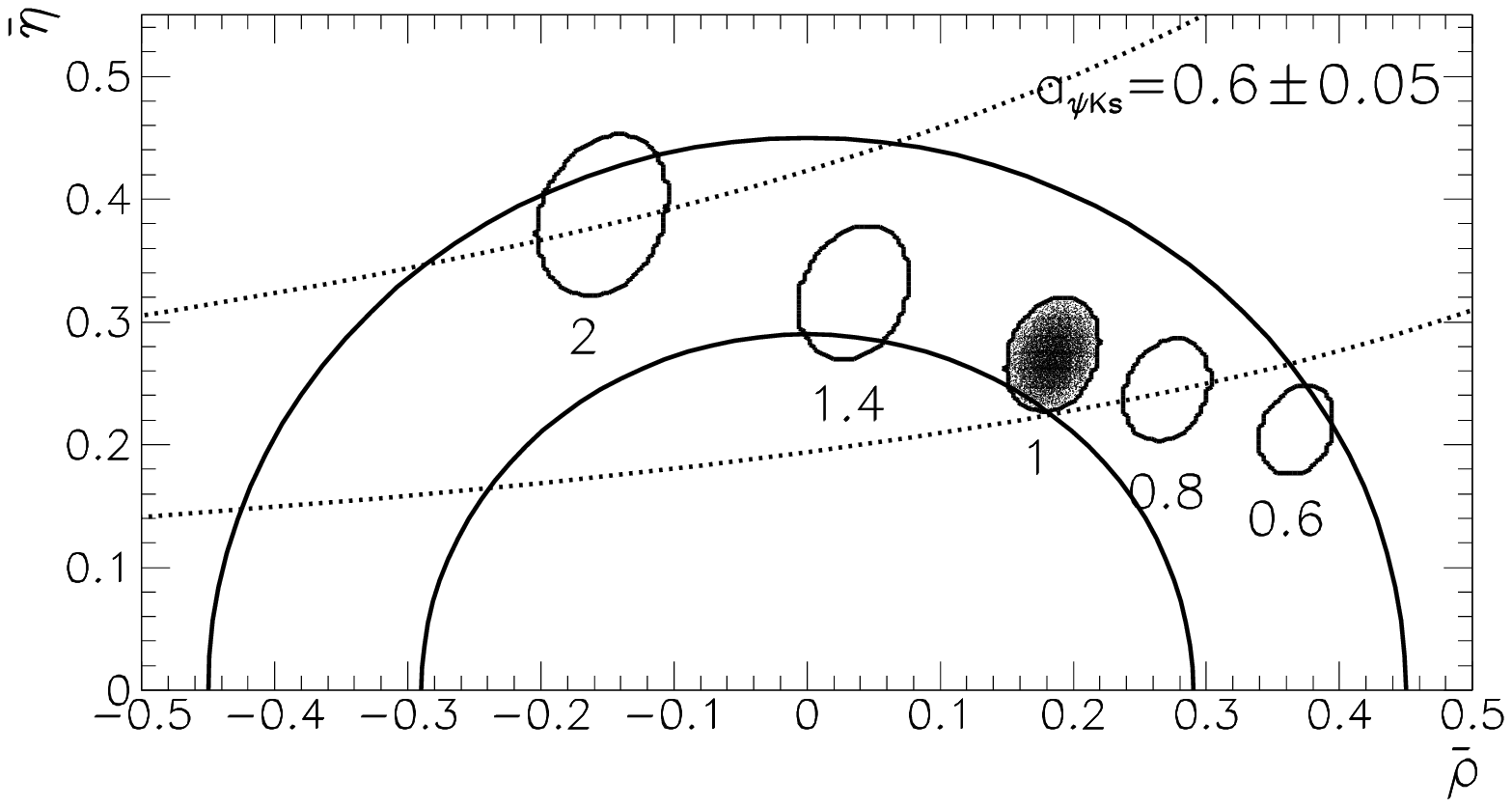,width=\linewidth}}\\
\mbox{\epsfig{file=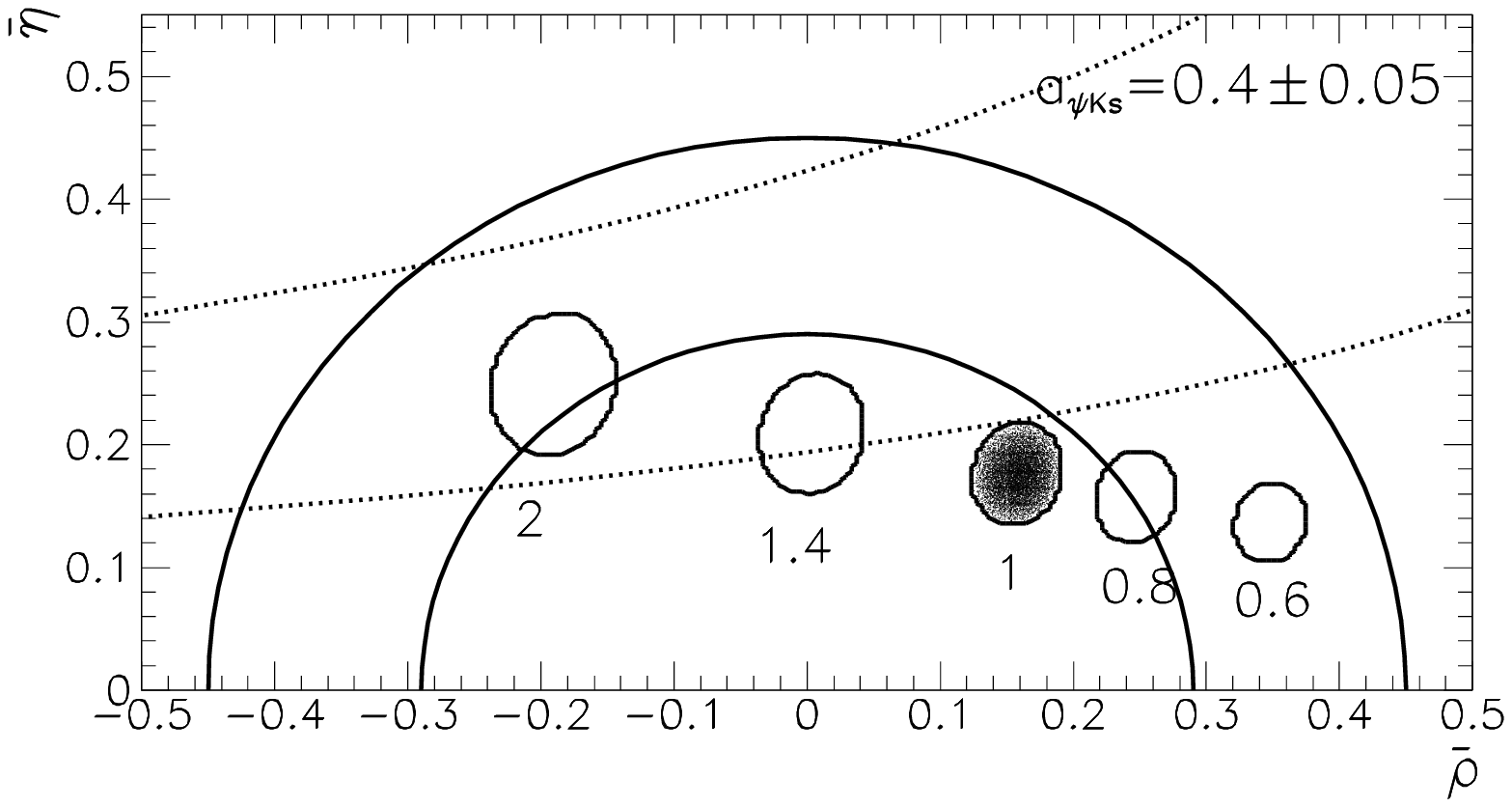,width=\linewidth}}\\
\end{tabular}
\caption{\protect Ranges of $(\bar\rho,\bar\eta)$ allowed in $1\sigma$ for 
  $\Delta M_s =(18.0\pm 0.5)/ps$, three values of $a_{\psi K_S}$ and
  different values of $R_{sd}$ (marked in the figures).  Black spots
  correspond to $R_{sd}=1$. Dotted lines show the constraint from the
  parameter $\varepsilon$, eq.~(\ref{eqn:epsk}), for
  $1+f_{\varepsilon}=1$.
\label{fig:bcrs4}}
\end{center}
\end{figure}

\subsubsection{Unitarity Triangle, \boldmath{$\sin2\beta$} and 
\boldmath{$\gamma$}}

A different version of strategy B is to construct the unitarity
triangle by means of the ratio $\Delta M_s/\Delta M_d$ and the
asymmetry $a_{\psi K_S}$. With $R_t$ given by (\ref{Rt}) the
parameters $\bar\varrho$ and $\bar\eta$ can be determined from the
formulae\footnote{This is valid for $-\pi/4<\beta<\pi/4$; for other
ranges of $\beta$ similar formulae can be obtained.}
\begin{eqnarray} 
&&\bar\eta={R_t\over2}\left(\sqrt{1+\sin2\beta}-\sqrt{1-\sin2\beta}\right)
\nonumber\\
&&\bar\rho=1-{R_t\over2}\left(\sqrt{1+\sin2\beta}+\sqrt{1-\sin2\beta}\right)
\phantom{aa}\label{eqn:barrhoeta}
\end{eqnarray}
(recall that in in GMFV models $\sin2\beta=a_{\psi K_S}$) obtained
directly from eqs.~(\ref{eqn:rtdef}) and (\ref{ss}).  These formulae
are equivalent to those presented in \cite{BUGAGOJASI,BUFL} but are
more elegant.

As an illustration we show in fig.~\ref{fig:bcrs4} the ranges of
$(\bar\varrho,\bar\eta)$ allowed by the hypothetical measurement
$\Delta M_s=(18.0\pm0.5)/ps$ for three values of $a_{\psi K_S}$ and
different values of $R_{sd}$. Solid ellipses correspond to 
$R_{sd}=1$ valid in particular in MFV models. We also show the
$R_b$-constraint, eq.~(\ref{eqn:Rb}), with $R_b=0.37\pm 0.08$ and as a
useful reference the $\varepsilon$-constraint (\ref{eqn:epsk}) with
$1+f_\varepsilon=1$ corresponding to the SM.

\begin{figure}[htbp]
\begin{center}
\epsfig{file=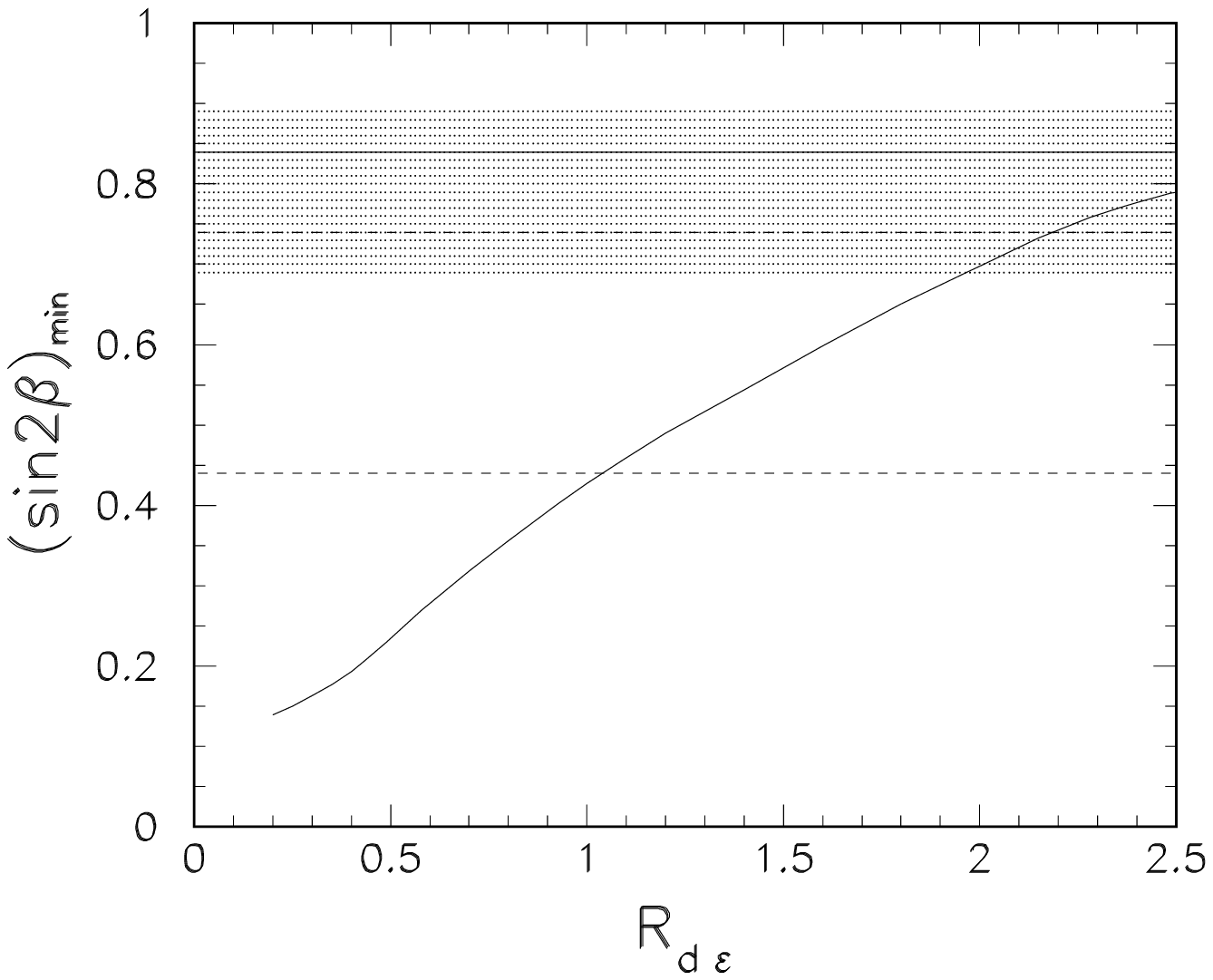,width=0.6\linewidth}
\end{center}
\caption{\protect Lower bound on $\sin2\beta$ in GMFV models as a function 
of the ratio $R_{d\varepsilon}$. Dashed lines show the 1-sigma BaBar result 
($\sin2\beta=0.59\pm0.15$), the horizontal solid line shows the lower limit 
{}from the Belle measurement ($\sin2\beta=0.99\pm0.15$) and the shaded area 
corresponds to the $\sin2\beta$ range presently allowed in GMFV models at 
$1\sigma$ by the official grand average (\ref{ga}).}
\label{fig:bcrs5}
\end{figure}

{}From fig.~\ref{fig:bcrs4} it is clear that for $a_{\psi K_S}=0.80\pm0.05$ 
the class of models giving $R_{sd}=1$ and $1+f_\varepsilon=1$ (which includes 
also the MFV models) is consistent with all constraints but only for 
$\sin2\beta$ in the lower part of the chosen range. There is also a room 
for contributions of new operators resulting in $R_{sd}\neq1$ provided 
$0.7\le R_{sd}\le 1.4$. As in strategy B they could be distinguished through 
the value of the angle $\gamma$. We also observe that no sizable contributions 
to $\varepsilon$ beyond the SM ones are required. For 
$a_{\psi K_S}=0.60\pm0.05$ models giving $R_{sd}=1$ are consistent with the 
$R_b$- and $\varepsilon$-constraints  for $\sin2\beta$ in the full chosen 
range. In this scenario models with $0.8\le R_{sd}\le 2.0$ and no sizable new 
contributions to $\varepsilon$ and models with $R_{sd}<0.8$ but with 
$1+f_\varepsilon >1.0$ are favoured. Again the measurement of $\gamma$ could 
distinguish between these possibilities. It is interesting to note that for 
$1.5\le R_{sd}\le 2.0$ it is possible to have $\gamma>90$ even for 
$1+f_\varepsilon=1$. Finally, for $a_{\psi K_S}=0.4$ models giving $R_{sd}=1$ 
are ruled out as they do not satisfy the $R_b$-constraint. In order to 
reconcile this constraint with $a_{\psi K_S}\approx0.4$, values of $R_{sd}$ 
substantially different from unity are required. Moreover in the case of 
$R_{sd}<1.0$ one has $\gamma\ll90^\circ$ but large new contributions to 
$\varepsilon$ leading to $1+f_\varepsilon>1.0$ and $R_{d\varepsilon}<1.0$ 
are mandatory. In contrast, if $R_{sd}\simgt2.0$ $\gamma$ can be much bigger 
than $90^\circ$ even without new contributions to $\varepsilon$ i.e. with
$1+f_\varepsilon\approx1$.

As in some scenarios discussed above $R_{d\varepsilon}$ must differ from 
unity, we show in fig.~\ref{fig:bcrs5} the dependence of the lower bound 
for $\sin2\beta$ on the value of $R_{d\varepsilon}$ together with 
the $1\sigma$ BaBar result and the
present official experimental $1\sigma$ band (\ref{ga}) for 
$a_{\psi K_S}=\sin2\beta$. It follows, that at present the ratio
$R_{d\varepsilon}$ would be constrained at $1\sigma$ by the BaBar result
to be less than 2.2 but the analogous limit following from the grand
average (\ref{ga}) is much higher. 
Fig.~\ref{fig:bcrs5} has been obtained for $R_{sd}=1.0$ but  
$(\sin2\beta)_{\rm min}$ depends only very weakly on the ratio 
$R_{sd}$: very similar curves are obtained also for $0.6\le R_{sd}\le 2.0$.

\subsection{Constraints on GMFV models from unitarity triangle}

In the preceding subsection we have presented two strategies which in
principle should allow to decide on the basis of experimental
measurements whether going beyond the MFV models is necessary, i.e. to
establish whether $R_{sd}\neq1$. In this section we want to explore
constraints and correlations imposed by the experimental data (present
and future) and the unitarity of the CKM matrix on the functions
$F_{tt}^i$. These constraints can be then effectively used to test the
specific GMFV models of new physics.

The first constraint follows from fitting the formula (\ref{eqn:xds})
to the measured (in the near future) value of $\Delta M_s$. This
determines $1+f_s$ (or $F_{tt}^s$):
\begin{eqnarray}
1+f_s=0.80 ~\left[{2.38\over S_0(x_t)}\right]
             \left[\frac{265~{\rm MeV}}{\sqrt{\hat B_{B_s}} F_{B_s}}\right]^2
             \left[\frac{0.55}{\eta_B}\right]
             \left[\frac{0.041}{|V_{ts}|}\right]^2
             \left[{\Delta M_s\over15/{\rm ps}}\right]
\label{eqn:fsp1cond}
\end{eqnarray}
This formula follows also by equating $R_t$ determined from $\Delta
M_d$ alone (eq.~(\ref{RT})) and $R_t$ determined from the ratio
$\Delta M_s/\Delta M_d$ (eq.~(\ref{Rt})). Scanning over uncertainties
in the ranges specified in table~\ref{tab:inputparams} and setting
$|V_{ts}|=|V_{cb}|$ gives
\begin{eqnarray}
0.52\left[{\Delta M_s\over15/ps}\right]<1+f_s<
1.29\left[{\Delta M_s\over15/ps}\right]
\label{eqn:1pfsbound}
\end{eqnarray}
(At present this gives of course $1+f_s>0.52$.)  It is worth
emphasizing that this bound is independent of the uncertainties of
$|V_{ub}/V_{cb}|$ as well as of any assumptions about possible new
physics contribution in the \KK system. Therefore the GMFV models not
respecting it are (will be) ruled out. We will see in sec. 4.2 that
the MSSM, for some values of its parameters, violates precisely this
bound.

Next, there are bounds on $R_t$ coming from the requirement that 
\begin{eqnarray}
1-R_b<R_t<1+R_b\label{eqn:rtrbbound}
\end{eqnarray}
which gives $0.54<R_t<1.46$. This can be used to constrain either
$1+f_d$ or $R_{sd}$ depending on how one determines $R_t$.  In the
first case one gets
\begin{eqnarray}
0.20 < 1+f_d < 4.24
\label{eqn:1pfdbound}
\end{eqnarray}
and in the second
\begin{eqnarray}
0.29 ~\left[{\Delta M_s\over15/ps}\right]< R_{sd} <
2.73  ~\left[{\Delta M_s\over15/ps}\right].
\label{eqn:rsdbound}
\end{eqnarray}

\begin{figure}[htbp]
\begin{center}
\epsfig{file=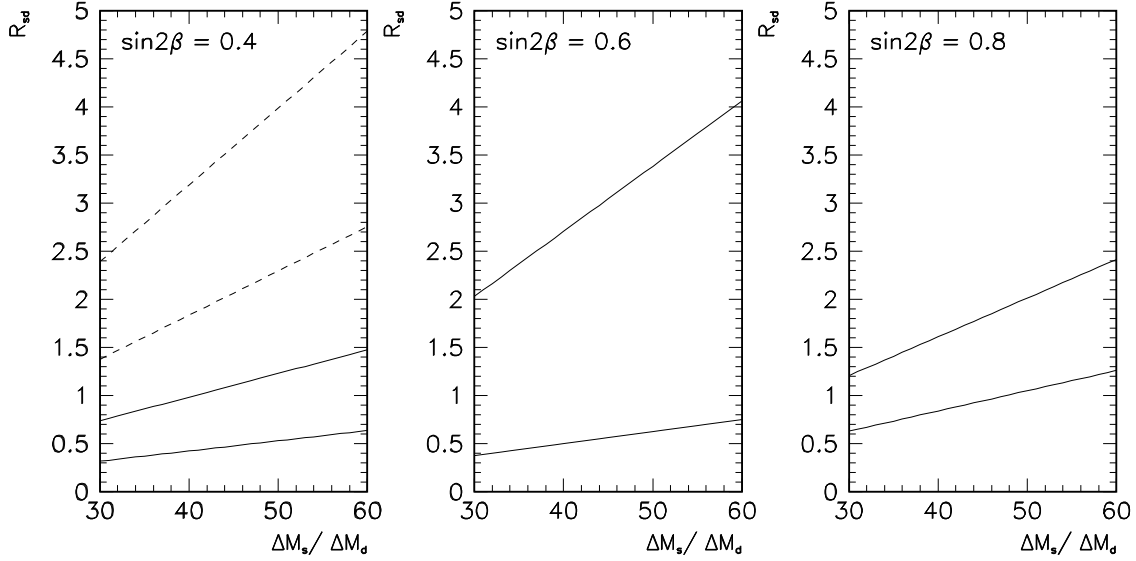,width=\linewidth}
\end{center}
\caption{\protect Allowed bands of $R_{sd}$ as a function of 
$\Delta M_s/\Delta M_d$ for different values of $\sin2\beta$.}
\label{fig:bcrs6}
\end{figure}

More stringent constraint on $R_{sd}$ (if $R_t$ is determined from
(\ref{Rt})) or $1+f_d$ (if $R_t$ is determined from (\ref{RT})) will
follow from $R_b$ combined with the information about $\sin2\beta$
obtained from future accurate measurements of the asymmetry $a_{\psi
K_{\rm S}}$. It is easy to see that for $\sin2\beta\simlt0.34$ there
are two allowed bands of $1+f_d$ corresponding to two possible
solutions\footnote{The band of allowed $R_t$ values splits into two
for $\sin2\beta\approx 0.56$ corresponding to $\sin\beta=R_b^{\rm
min}$.  Additional uncertainties in translating $R_t$ into $1+f_d$ (or
$R_{sd}$) result in lowering the value of $\sin2\beta$ below which the
two allowed bands of $1+f_d$ (or $R_{sd}$) appear.} for $R_t$:
\be
R_t=\cos\beta\mp\sqrt{R_b^2-\sin^2\beta}
\ee
(the solutions with $-(+)$ correspond to smaller (larger) values of
the angle $\gamma$). For larger values of $\sin2\beta$ the two bands
overlap which means that there is only one allowed range of $1+f_d$
(or $R_{sd}$).  For example scanning over uncertainties one obtains:
$0.21<1+f_d<0.78$ or $0.84<1+f_d<4.15$ for $\sin2\beta=0.2$,
$0.23<1+f_d<3.85$ for $\sin2\beta=0.4$, $0.26<1+f_d<3.27$ for
$\sin2\beta=0.6$ and $0.44<1+f_d<1.98$ for $\sin2\beta=0.8$. The
corresponding allowed ranges of $R_{sd}$ are shown in
fig.~\ref{fig:bcrs6} for different values of $\sin2\beta$ as functions
of the measured values of $\Delta M_s/\Delta M_d$.

\begin{figure}[htbp]
\begin{center}
\epsfig{file=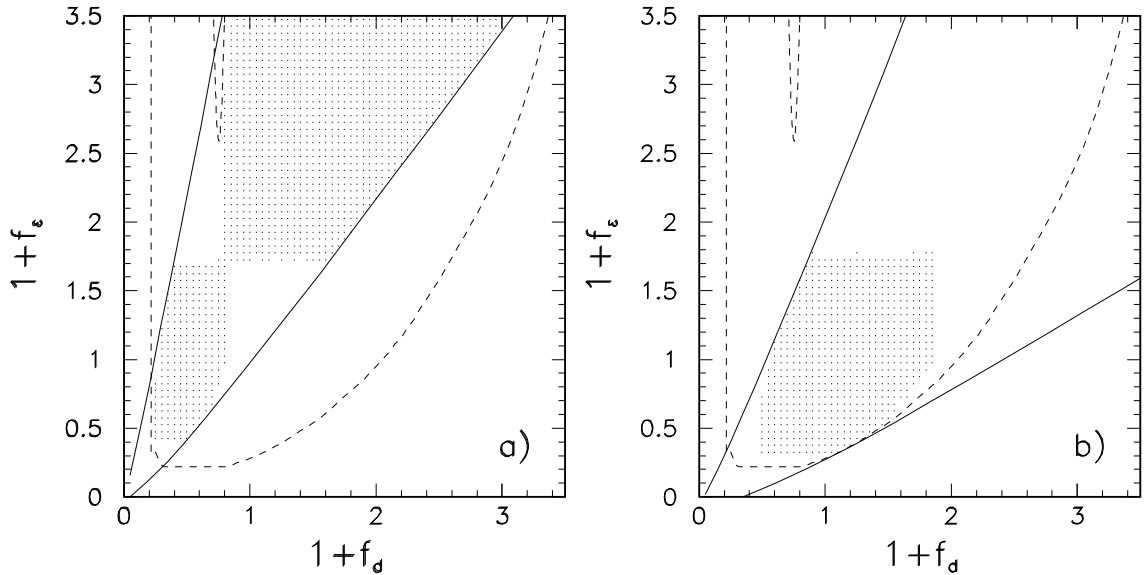,width=\linewidth}
\end{center}
\caption{\protect Allowed ranges of $1+f_d$ and $1+f_\varepsilon$. 
  $\Delta M_d$, $\varepsilon$ and $R_b$ allow the region delimited by
  the dashed lines. Regions between the solid lines are allowed by
  $\Delta M_d$, $\varepsilon$ and $\sin2\beta=0.4$ (panel a) and
  $\sin2\beta=0.8$ (panel b). Dotted regions are allowed by $\Delta
  M_d$, $\varepsilon$ and $R_b$ for $\sin2\beta=0.4$ and $0.8$ in
  panels a) and b), respectively.}
\label{fig:bcrs7}
\end{figure}

Further constraints correlating $1+f_\varepsilon$ with $1+f_d$ can be 
obtained by using the experimental information about the parameter 
$\varepsilon$. To this end, with a range of $R_t$ determined from $1+f_d$ 
and $\Delta M_d$ by scanning over the relevant uncertainties, one checks 
whether there exist values of $\bar\rho$ and $\bar\eta$ satisfying 
eqs.~(\ref{eqn:rtdef}) and (\ref{eqn:epsk}) for 
a given value of $1+f_\varepsilon$. It is easy to see that from $\Delta M_d$ 
and $\varepsilon$ alone only a very weak lower bound  on $|1+f_\varepsilon|$ 
can exist for $1+f_d\simgt1.8$. As a next step, one can impose the 
constraint from $R_b$ (eq.~(\ref{eqn:Rb})).
The resulting allowed range in the plane $(1+f_d,~1+f_\varepsilon)$ is shown
in fig.~\ref{fig:bcrs7}a,b by the dashed lines (their vertical parts
correspond to the lower bound (\ref{eqn:1pfdbound})). For $1+f_d$ in the 
range (\ref{eqn:1pfdbound}) no upper limit on $|1+f_\varepsilon|$ from
$\Delta M_d$, $\varepsilon$ and $R_b$ exists (except for a very narrow 
range $0.7<1+f_d<0.85$). This is because for almost all values of $1+f_d$ 
satisfying (\ref{eqn:1pfdbound}) the range of possible values of $R_t$ is 
such that it is possible to satisfy the constraint from $R_b$ with
$\bar\eta=0$ which in turn allows to suppress arbitrarily large values of
$|1+f_\varepsilon|$ and to satisfy the eq.~(\ref{eqn:epsk}).

\begin{figure}[htbp]
\begin{center}
\epsfig{file=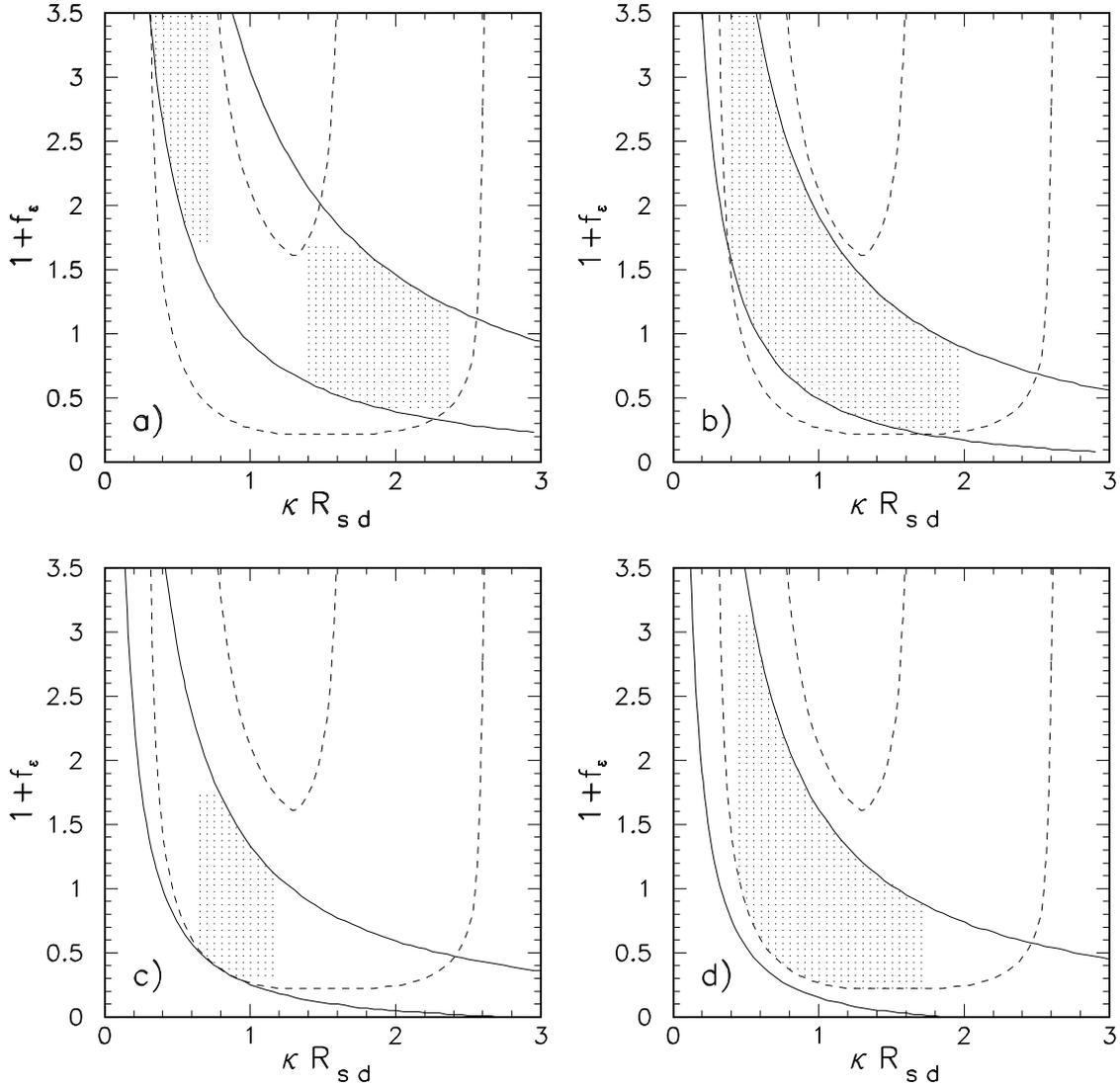,width=\linewidth}
\end{center}
\caption{\protect Allowed ranges of $\kappa R_{sd}$ and $1+f_\varepsilon$
  where $\kappa\equiv{30\over\Delta M_s/\Delta M_d}$. $\Delta M_s/\Delta M_d$,
  $\varepsilon$ and $R_b$ allow the region delimited by the dashed
  lines. Regions between the solid lines are allowed 
  by $\Delta M_s/\Delta M_d$,
  $\varepsilon$ and $\sin2\beta=0.4$ (panel a), $\sin2\beta=0.6$
  (panel b), $\sin2\beta=0.8$ (panel c) and $\sin2\beta=0.79\pm10$
  (panel d). Dotted regions are allowed by $\Delta M_s/\Delta M_d$,
  $\varepsilon$, $R_b$ and the corresponding value of $\sin2\beta$.}
\label{fig:bcrs8}
\end{figure}

It is also interesting to assume that the asymmetry $a_{\psi K_{\rm S}}$
(i.e. $\sin2\beta$ in GMFV models) is measured with sufficient accuracy and to
correlate $1+f_\varepsilon$ with $1+f_d$ by using the experimental information
about $\Delta M_d$ and $\varepsilon$ for fixed values of $\sin2\beta$. To 
this end, for $R_t$ given by eq.~(\ref{RT}) one determines the
parameters $\bar\rho$ and $\bar\eta$ from the formulae (\ref{eqn:barrhoeta})
and checks whether eq.~(\ref{eqn:epsk}) can be satisfied. Constraints 
on the plane $(1+f_d,~1+f_\varepsilon)$ obtained in this way are shown in 
fig.~\ref{fig:bcrs7}a for $\sin2\beta=0.4$ and in fig.~\ref{fig:bcrs7}b for 
$\sin2\beta=0.8$ by the solid lines. It should be stressed that it is 
the constraint from the $a_{\psi K_{\rm S}}$ asymmetry 
which eliminates solutions with $\bar\eta\approx0$ thus providing, for fixed 
$1+f_d$, the upper bound on $1+f_\varepsilon$. In agreement with the bound 
(\ref{bound}) we observe that for $\sin2\beta=0.4$, 
$R_{d\varepsilon}\equiv(1+f_d)/(1+f_\varepsilon)$ has to deviate from unity 
while for $\sin2\beta=0.8$ points corresponding to $R_{d\varepsilon}=1$ lie 
within the allowed region. 

Finally, one can impose also the constraints from $R_b$  (\ref{eqn:Rb}). 
The allowed ranges in the $(1+f_d,~1+f_\varepsilon)$ planes are dotted
in figs.~\ref{fig:bcrs7}a,b. They are not simply given by the intersection
of the regions allowed respectively by $(\Delta M_d, ~\varepsilon, ~R_b)$
and $(\Delta M_d, ~\varepsilon, ~\sin2\beta)$ because the same point in
the $(1+f_d,~1+f_\varepsilon)$ plane may require different $\bar\rho$ and
$\bar\eta$ to be compatible with the two above sets of experimental data.

For a fixed value of the ratio $\Delta M_s/\Delta M_d$ the same
analysis can be also repeated for the $(R_{sd},~1+f_\varepsilon)$
plane.  However, since the value of $\Delta M_s/\Delta M_d$ serves
only to determine $R_t$ from eq.~(\ref{Rt}) we show instead in
fig.~\ref{fig:bcrs8} the allowed ranges in the plane $(\kappa R_{sd},
1+f_\varepsilon)$ where $\kappa\equiv{30\over\Delta M_s/\Delta M_d}$. 
In this figure, panels a)-c) correspond to $\sin2\beta=0.4,~0.6,~0.8$ 
respectively. Panel d) corresponds to the averaged
experimental value $0.80\pm0.11$.  Because of the absolute constraint
(\ref{eqn:fsp1cond}), $R_{sd}\propto1/(1+f_d)$ and qualitatively
figs.~\ref{fig:bcrs8}a and \ref{fig:bcrs8}b are obtained from
figs.~\ref{fig:bcrs7}a and \ref{fig:bcrs7}b, respectively by
subjecting the $x$-axis to the transformation $x\rightarrow1/x$.
Quantitatively however, the bounds on GMFV models provided by
fig.~\ref{fig:bcrs8} are tighter as they are independent of the poorly
known parameter $F_{B_d}\sqrt{\hat B_{B_d}}$.

\section{General formulae for \boldmath{$F_{tt}^d$}, \boldmath{$F_{tt}^s$}
 and \boldmath{$F_{tt}^\varepsilon$} in GMFV Models}
\setcounter{equation}{0}

Using general formulae (7.27)--(7.32) in \cite{BUJAUR} it is easy to 
express the functions $F_{tt}^d$, $F_{tt}^s$ and $F_{tt}^\varepsilon$ in 
terms of the Wilson coefficients at the scale at which the effective 
Hamiltonian is generated, the relevant QCD renormalization group factors 
$\eta$ and the non-perturbative $B_i$ factors. Suppressing the superscripts 
$d$, $s$ and $\varepsilon$ for a moment we find
\begin{eqnarray}\label{hds2}
F_{tt} &=& \left[S_0(x_t)+ \frac{1}{4 r}C_{\rm new}^{\rm VLL}(\mu_t)\right]
\nonumber\\
&&+{1\over4r}C^{\rm VRR}_1(\mu_t)+
\bar P_1^{\rm LR} C^{\rm LR}_1(\mu_t) 
+\bar P_2^{\rm LR} C^{\rm LR}_2(\mu_t) \nonumber\\
&&+\bar P_1^{\rm SLL}\left[C^{\rm SLL}_1(\mu_t)+C^{\rm SRR}_1(\mu_t)\right]
+\bar P_2^{\rm SLL}\left[C^{\rm SLL}_2(\mu_t)+C^{\rm SRR}_2(\mu_t)\right]
\end{eqnarray}
where $r=0.985$ \cite{BUJAWE} describes the QCD corrections to
$S_0(x_t)$ in the SM. This factor is present because we have factored
out $\eta_B$ in (\ref{eqn:xds}) and $\eta_2$ in the analogous formula
for $\varepsilon$. The first line of eq. (\ref{hds2}) contributes to 
$f_{\rm uni}$ in eq. (\ref{uni}) and is therefore present also in the 
MFV models. The remaining lines are characteristic for the GMFV models 
and contribute to the $\tilde f_i$ in eq. (\ref{uni}). Thus, different 
values of the functions $F_{tt}^d$, $F_{tt}^s$ and $F_{tt}^\varepsilon$ 
originate from generally different values of the Wilson coefficients 
$C^a_i(\mu_t)$ and factors $\bar P^a_i$ pertinent to $\Delta M_d$, 
$\Delta M_s$ and $\varepsilon$:
\begin{equation}\label{pab}
\bar P^a_i=\left\{\begin{array}{ll}
{P_i^a}/{(4\eta_B\hat B_{B_{d,s}})} & (\Delta M_{d,s}) \\
{P_i^a}/{(4\eta_2\hat B_K)} & (\varepsilon)
\end{array}\right.~,
\end{equation}
where $\hat B_{B_{d,s}}$ and $\hat B_K$ are the relevant non-perturbative 
parameters related to the matrix elements of $Q_1^{\rm VLL}$.

In the case of $F_{tt}^{d,s}$ the coefficients $P^a_i$ are given by
\be
P_1^{\rm LR} =-\frac{1}{2} \left[\eta_{11} (\mu_b)\right]_{\rm LR}
               \left[B^{\rm LR}_1(\mu_b)\right]_{\rm eff}
+\frac{3}{4} \left[\eta_{21} (\mu_b)\right]_{\rm LR}
               \left[B^{\rm LR}_2(\mu_b)\right]_{\rm eff},
\ee
\be
P_2^{\rm LR}=-\frac{1}{2} \left[\eta_{12} (\mu_b)\right]_{\rm LR}
               \left[B^{\rm LR}_1(\mu_b)\right]_{\rm eff}
+\frac{3}{4} \left[\eta_{22} (\mu_b)\right]_{\rm LR}
               \left[B^{\rm LR}_2(\mu_b)\right]_{\rm eff},
\ee

\be
P_1^{\rm SLL} =-\frac{5}{8} \left[\eta_{11} (\mu_b)\right]_{\rm SLL}
               \left[B^{\rm SLL}_1(\mu_b)\right]_{\rm eff}
-\frac{3}{2} \left[\eta_{21} (\mu_b)\right]_{\rm SLL}
               \left[B^{\rm SLL}_2(\mu_b)\right]_{\rm eff},
\ee
\be
P_2^{\rm SLL}=-\frac{5}{8} \left[\eta_{12} (\mu_b)\right]_{\rm SLL}
               \left[B^{\rm SLL}_1(\mu_b)\right]_{\rm eff}
-\frac{3}{2} \left[\eta_{22} (\mu_b)\right]_{\rm SLL}
               \left[B^{\rm SLL}_2(\mu_b)\right]_{\rm eff}.
\ee
with the effective parameters $\left[B^a_i(\mu_b)\right]_{\rm eff}$ defined by
\begin{eqnarray}
&&\left[B^a_i(\mu_b)\right]_{\rm eff}\equiv
\left({M_{B_q}\over m_b(\mu_b)+m_q(\mu_b)}\right)^2 B^a_i(\mu_b)
\phantom{aa}\nonumber\\
&&\phantom{aaaa}=1.44
\left[{4.4~{\rm GeV}\over m_b(\mu_b)+m_q(\mu_b)}\right]^2
\left[{M_{B_q}\over5.28~{\rm GeV}}\right]^2 B^a_i(\mu_b),
\label{eff}
\end{eqnarray}
where $B^a_i(\mu_b)$ are related to the hadronic matrix elements
$\langle\bar B_q^0|Q_i|B_q^0\rangle$. The QCD factors
$\left[\eta_{ij} (\mu_b)\right]_a$ are given in \cite{BUJAUR}.

In the case of $F_{tt}^\varepsilon$ the QCD factors 
$\left[\eta_{ij}(\mu_b)\right]_a$ are replaced with 
$\left[\eta_{ij}(\mu_L)\right]_a$ where $\mu_L=2$~GeV 
and the corresponding effective parameters
$\left[B^a_i(\mu_L)\right]_{\rm eff}$ are defined by
\be
\label{Balat}
\left[B^a_i(\mu_L)\right]_{\rm eff}\equiv
\left(\frac{m_K}{m_s(\mu_L)+m_d(\mu_L)}\right)^2 B^a_i(\mu_L)=18.75
\left[\frac{115~{\rm MeV}}{m_s(\mu_L)+m_d(\mu_L)}\right]^2B^a_i(\mu_L).
\ee

The NLO QCD factors $\eta_{ij}(\mu_b)$ and $\eta_{ij}(2$~GeV),
relevant for $\Delta M_{d,s}$ and $\varepsilon$ respectively, have
been evaluated in \cite{BUJAUR}. For completeness we give in
table~\ref{table1} their numerical values for different scales
(denoted here by $\mu_{\rm NP}$) at which the new operators are
generated.

\begin{table}
\begin{center}
\begin{tabular}{|l||c|c|c||c|c|c|}
\hline
\TT&\multicolumn{3}{c||}{$\mu_b=4.2$ GeV}&
    \multicolumn{3}{c|}{$\mu_L=2$ GeV}\\
\hline
\TT$\mu_{\rm NP}$& $\mu_t$ & 500 GeV & 1 TeV & $\mu_t$ & 500 GeV & 1 TeV\\
\hline
$\TT\left[\eta\right]_{\rm  VLL}$&0.838&0.809&0.793&0.787&0.759&0.744\\[2mm]
$\left[\eta_{11}\right]_{\rm LR}$&0.919&0.907&0.902&0.906&0.900&0.898\\[2mm]
$\left[\eta_{12}\right]_{\rm LR}$&$-$0.043&$-$0.054&$-$0.060&$-$0.089&
$-$0.107&$-$0.118\\[2mm]
$\left[\eta_{21}\right]_{\rm LR}$&$-$0.919&$-$1.190&$-$1.360&$-$1.548&
$-$1.923&$-$2.159\\[2mm]
$\left[\eta_{22}\right]_{\rm  LR}$&2.303&2.701&2.951&3.227&3.785&4.136\\[2mm]
$\left[\eta_{11}\right]_{\rm SLL}$&1.676&1.846&1.949&2.063&2.272&2.398\\[2mm]
$\left[\eta_{12}\right]_{\rm SLL}$&2.049&2.470&2.715&2.970&3.441&3.717\\[2mm]
$\left[\eta_{21}\right]_{\rm SLL}$&$-$0.007&$-$0.008&$-$0.009&$-$0.009&
$-$0.011&$-$0.012\\[2mm]
$\left[\eta_{22} \right]_{\rm SLL}$&0.540&0.480&0.449&0.414&0.366&0.341\\[2mm]
\hline
\end{tabular}
\caption{Numerical values for the $\eta$-factors for the \BB and \KK
mixing for $\alpha_s^{(5)}(M_Z)=0.118$ and different values of the
scale $\mu_{\rm NP}$ at which the New Physics is integrated out.}
\label{table1}
\end{center}
\end{table}

The parameters $B_i$ for the \KK mixing are known from lattice 
calculations  \cite{CET,latt_lit}. For the values corresponding to
$\mu_t$ in the table \ref{table1} and in  the NDR scheme one finds 
\cite{BUJAUR}:
\begin{eqnarray}
\left.\matrix{P_1^{\rm LR} =-36.1, & P_2^{\rm LR}= 59.3,\cr
P_1^{\rm SLL} =-18.1, & P_2^{\rm SLL}=-32.2,}\right\} 
\qquad {\rm for} \qquad \mu=2~{\rm GeV}.
\end{eqnarray}
The large values of these coefficients originate in the strong enhancement 
of the QCD factors $\eta_{ij}$ for the LR and SLL (SRR) operators and in the 
chiral enhancement of their
matrix elements seen in eq.~(\ref{Balat}). Consequently even small new physics 
contributions to $C^{\rm LR}_i(\mu_t)$ and $C^{\rm SLL}_i(\mu_t)$ 
can play an important role in the phenomenology \cite{CET,BAMAZH}.

In the case of the \BB mixing the chiral enhancement of the hadronic
matrix elements of the LR and SLL operators is absent. Moreover, the
QCD factors $\eta_{ij}$ are smaller than in the case of the \KK
mixing. Consequently the coefficients $P_i^{\rm LR}$ and $P_i^{\rm
SLL}$ are smaller in this case but can be still important.  As lattice
results are not yet available for the hadronic matrix elements of the
LR and SLL operators in the $B$ system \cite{FLLISA} we will set in
this case $B_i^a(\mu_b)=1$.  Taking $M_B=5.28$ GeV, $\mu_b=4.2$ GeV,
$m_b(\mu_b)+m_q(\mu_b)=4.2$ GeV and $\alpha_s(M_Z)=0.118$ one finds
\begin{eqnarray}
\left.\matrix{P_1^{\rm LR} =-1.65, & P_2^{\rm LR}= 2.51, \cr
P_1^{\rm SLL} =-1.49, &P_2^{\rm SLL}=-3.01,}\right\}\qquad {\rm for} 
\qquad \mu_b=4.2~{\rm GeV} .
\end{eqnarray}
Since we took $\mu_b=4.2$ GeV instead of $4.4$ GeV, these numbers differ
slightly from those given in ref.~\cite{BUJAUR}.
Finally in order to calculate $\bar P_i^a$ in (\ref{pab}) we will use 
$\eta_B=0.55$, $\eta_2=0.57$ and \cite{FLLISA}
\be
\hat B_K=0.85\pm0.15, \qquad \hat B_{B_{d,s}}=1.30\pm 0.18~.
\ee 

\section{\boldmath{$F_{tt}^d$}, \boldmath{$F_{tt}^s$} and 
\boldmath{$F_{tt}^\varepsilon$} in realistic GMVF Models}

\setcounter{equation}{0}

\subsection{2HDM(II) with large $\tan\bar\beta$}

To see what values of $F_{tt}^d$, $F_{tt}^s$ and $F_{tt}^\varepsilon$
can be realized in realistic GMFV models we consider here two
extensions of the SM. The first is the 2HDM(II) which introduces three
neutral physical scalars ($h^0$, $H^0$ and $A^0$) and a charged
physical scalars $H^\pm$. At one loop only the charged scalars are
relevant for the box diagrams contributing to \KK and $\bar B^0$-$B^0$
mixing amplitudes. Using the compact notation of ref. \cite{ROS} the
tree level couplings of $H^+_k\equiv(H^+,G^+)$ (where $G^\pm$ is the
would-be Goldstone boson) read
\begin{eqnarray}
L_{\rm int}=H^+_k\bar u_AV_{AI}(a^{AIk}_LP_L+a^{AIk}_RP_R)d_I+{\rm Hc.}
\end{eqnarray}
where
\begin{eqnarray}
a^{AIk}_L=
{e\over\sqrt2s_W}{m_{u_A}\over M_W}\times\left\{\matrix{
\cot\bar\beta ~~~~~{\rm for} ~~k=1\cr
\phantom{aa} 1 \phantom{aaa} ~~~{\rm for} ~~k=2
}\right.
\label{eqn:Lcouplings}
\end{eqnarray}
\begin{eqnarray}
a^{AIk}_R=
{e\over\sqrt2s_W}{m_{d_I}\over M_W}\times\left\{\matrix{
\tan\bar\beta  ~~~~~{\rm for} ~~k=1\cr
\phantom{a} -1\phantom{aaa} ~~{\rm for} ~~k=2}\right.
\label{eqn:Rcouplings}
\end{eqnarray}
(Recall that in our notation $d_3\equiv b$, $d_2\equiv s$, $d_1\equiv
d$ and, analogously, $u_3\equiv t$, $u_2\equiv c$, $u_1\equiv u$.)
Contribution of $H_k^\pm$ to the Wilson coefficients $C_i$ of the
operators responsible for the transition $d_I\bar d_J\rightarrow\bar
d_I d_J$ in eq.~(\ref{heff}) can be easily expressed in terms of the
coefficients $a^{AIk}_L$ and $a^{AIk}_R$. Diagrams with one $W^\pm$
and one $H^\pm$ give\footnote{The contribution of $G^\pm$ to
$C^{VLL}_1$ is already taken into account in the function
$S_0(x_t)$. Masses of the $u$ and $c$ quarks are neglected.}:
\begin{eqnarray}
&&G^2_FM^2_W\delta^{(+)}C^{\rm VLL}_1(\mu_{\rm NP})
=-{e^2\over2s^2_W}a^{tJ1}_La^{tI1}_L m^2_t
D_0(M_W,M_{H^+},m_t,m_t)\phantom{aa}\nonumber\\
&&G^2_FM^2_W\delta^{(+)}C^{\rm LR}_2(\mu_{\rm NP})=
-{e^2\over2s^2_W}\sum_{k=1}^2a^{tJk}_R a^{tIk}_R\left[
D_2(M_W,M_{H^+_k},m_t,m_t)\right.\nonumber\\
&&\phantom{aaaaaaaaaaaaaaaaa}
\left.-2D_2(M_W,M_{H^+_k},m_t,0)+D_2(M_W,M_{H^+_k},0,0)
\right]\phantom{aa}\label{eqn:WHcontr}
\end{eqnarray}
where the four-point functions $D_0$ and $D_2$ are defined in the
Appendix.  Diagrams with two $H_k^\pm$ give\footnote{In the sum over
$k$ and $l$ in the expression for $\delta^{(+)}C^{\rm VLL}_1$ the
contribution of $G^\pm G^\mp$ is excluded. It is taken into account in
the function $S_0(x_t)$.}:
\begin{eqnarray}
&&G^2_FM^2_W\delta^{(+)}C^{\rm VLL}_1(\mu_{\rm NP}) = {1\over8}\sum_{k,l}
a^{tJl}_L a^{tIk}_L a^{tJk}_L a^{tIl}_L 
D_2(M_{H^+_l},M_{H^+_k},m_t,m_t)\phantom{aa}\nonumber\\
&&G^2_FM^2_W\delta^{(+)}C^{\rm VRR}_1(\mu_{\rm NP}) = {1\over8}\sum_{k,l}^2
a^{tJl}_R a^{tIk}_R a^{tJk}_R a^{tIl}_R \left[
D_2(M_{H^+_l},M_{H^+_k},m_t,m_t)\right.\nonumber\\
&&\phantom{aaaaaaaaaaaaaaaaaaa}
\left.-2D_2(M_{H^+_l},M_{H^+_k},m_t,0)+D_2(M_{H^+_l},M_{H^+_k},0,0)
\right]\phantom{aa}\nonumber\\
&&G^2_FM^2_W\delta^{(+)}C^{\rm LR}_1(\mu_{\rm NP}) = {1\over4}\sum_{k,l}^2
a^{tJl}_L a^{tIk}_L a^{tJk}_R a^{tIl}_R 
D_2(M_{H^+_l},M_{H^+_k},m_t,m_t)\phantom{aa}\nonumber\\
&&G^2_FM^2_W\delta^{(+)}C^{\rm SLL}_1(\mu_{\rm NP}) = {1\over2}\sum_{k,l}^2
a^{tJl}_R a^{tIk}_L a^{tJk}_R a^{tIl}_L m_t^2
D_0(M_{H^+_l},M_{H^+_k},m_t,m_t)\phantom{aa}\nonumber\\
&&G^2_FM^2_W\delta^{(+)}C^{\rm SRR}_1(\mu_{\rm NP}) = {1\over2}\sum_{k,l}^2
a^{tJl}_L a^{tIk}_R a^{tJk}_L a^{tIl}_R m_t^2
D_0(M_{H^+_l},M_{H^+_k},m_t,m_t)\phantom{aa}\nonumber\\
&&G^2_FM^2_W\delta^{(+)}C^{\rm LR}_2(\mu_{\rm NP}) = \sum_{k,l}^2
a^{tJl}_R a^{tIk}_L a^{tJk}_L a^{tIl}_R m_t^2
D_0(M_{H^+_l},M_{H^+_k},m_t,m_t)\phantom{aa}
\label{eqn:HHcontr}
\end{eqnarray}
At one loop there are no contributions to the Wilson coefficients of
the tensor operators $Q_2^{\rm SLL}$ and $Q_2^{\rm SRR}$. In the
computations we take $\mu_{\rm NP}=M_{H^+}$ and apply the formulae
given in Appendix C of ref. \cite{BUJAUR}. As our calculation of the
Wilson coefficients at $\mu_{\rm NP}$ does not include ${\cal
O}(\alpha_s)$ corrections and the relevant matrix elements of LR and
SLL operators in the $B$-system have been evaluated using the vacuum
insertion method, there are inevitably unphysical scale and
renormalization scheme dependences present in our final results. We
expect that these dependences are small at scales ${\cal O}(\mu_{\rm
NP})$ as the strong coupling $\alpha_s(\mu_{\rm NP})$ is small. They
could turn out to be more important at $\mu=\mu_b$ where $\alpha_s$ is
bigger.  Consequently the evaluation of the hadronic matrix elements
of the LR and SLL operators relevant for \BB mixing in the NDR scheme
is very desirable. This would not only remove the unphysical
dependences in question but would also give the actual values of the
relevant matrix elements in QCD. Still we believe that our calculation
captures the correct size of the dominant new physics effects.

For large $\tan\bar\beta$ and $M_{H^+}\approx m_t$ the leading terms
of the above contributions to the Wilson coefficients $C_i$ are of the
order ($e^4/(32s^4_WM^2_W)=G_F^2M^2_W$):
\begin{eqnarray}
\delta^{(+)}C^{\rm VLL}_1\sim{4\over3}\cot^2\bar\beta ,\phantom{aa}
\delta^{(+)}C^{\rm LR}_2
\sim-{8\over3}{m_{d_I}m_{d_J}\over m^2_t}\tan^2\bar\beta
\label{eqn:WHorder}
\end{eqnarray}
for diagrams with $W^\pm H^\mp$, and
\begin{eqnarray}
\delta^{(+)}C^{\rm VLL}_1\sim
{1\over3}{m^2_t\over M^2_W}\cot^2\bar\beta  ,\phantom{aa}
\delta^{(+)}C^{\rm VRR}_1\sim{1\over3}{m^2_{d_I}m^2_{d_J}\over M^2_Wm^2_t}
\tan^4\bar\beta  ,\phantom{aa}
\delta^{(+)}C^{\rm LR}_1\sim 0\nonumber\\
\delta^{(+)}C^{\rm SLL}_1\sim 0 ,\phantom{aa}
\delta^{(+)}C^{\rm SRR}_1\sim 0 ,\phantom{aa}
\delta^{(+)}C^{\rm LR}_2
\sim-{4\over3}{m_{d_I}m_{d_J}\over M^2_W}\tan^2\bar\beta
\phantom{aaa}
\label{eqn:HHorder}
\end{eqnarray}
for diagrams with $H^\pm H^\mp$. It is clear that for large
$\tan\bar\beta$ the biggest contribution appears in
$\delta^{(+)}C^{\rm LR}_2$. It is of the opposite sign than the
contribution of the $tW^\pm$ box diagram and can be significant only
for the $\bar B^0_s$-$B^0_s$ transition amplitude for which it is of
the order
\begin{eqnarray}
\delta^{(+)}C^{\rm LR}_2\approx
-{2m_s(\mu_t)m_b(\mu_t)\over M^2_W}\tan^2\bar\beta
\approx-0.14\times\left({\tan\bar\beta\over50}\right)^2
\label{eqn:BBorder}
\end{eqnarray}
where we have used $m_b(\mu_t)\approx3$ GeV and $m_s(\mu_t)\approx61$
MeV.  Compared to the estimate of eq.~(\ref{eqn:BBorder}), similar
contributions to $\delta^{(+)}C^{LR}_2$ for $\bar B^0_d$-$B^0_d$ and
\KK transitions are suppressed by factors $m_d/m_s$ and $m_d/m_b$,
respectively. As discussed in Sec. 3 all these contributions are further
enhanced compared to the standard ones by the QCD renormalization
effects and, in the case of the \KK transition, also by the chiral
enhancement of the corresponding matrix element. As a result, in the
case of $\bar B^0_s$-$B^0_s$ mixing the contribution of the $Q^{\rm
LR}_2$ operator can compete with the contribution of the standard
$Q^{\rm VLL}_1$ one for light charged Higgs boson and large values of
$\tan\bar\beta$.  To demonstrate it we plot in fig.~\ref{fig:bcrs9}
the value of $1+f_s$ in 2HDM(II) for different values of its
parameters. Fig.~\ref{fig:bcrs9}a shows that a significant decrease of
$1+f_s$ below unity is possible only for $M_{H^+}\sim100-200$ GeV and
$\tan\bar\beta$ close to its upper limit following from the
requirement of perturbativity of the bottom quark Yukawa coupling. The
corresponding effects in $1+f_d$ and $1+f_\varepsilon$ are negligible.
The increase of $1+f_s$ above unity seen in fig.~\ref{fig:bcrs9}a for
$\tan\bar\beta<10$ reflects a well known universal contribution of the
box diagrams to the Wilson coefficient of the $Q^{\rm VLL}_1$ operator
which gives $1+f_s\approx1+f_d\approx1+f_\varepsilon>1$.

\begin{figure}[htbp]
\begin{center}
\begin{tabular}{p{0.48\linewidth}p{0.48\linewidth}}
\mbox{\epsfig{file=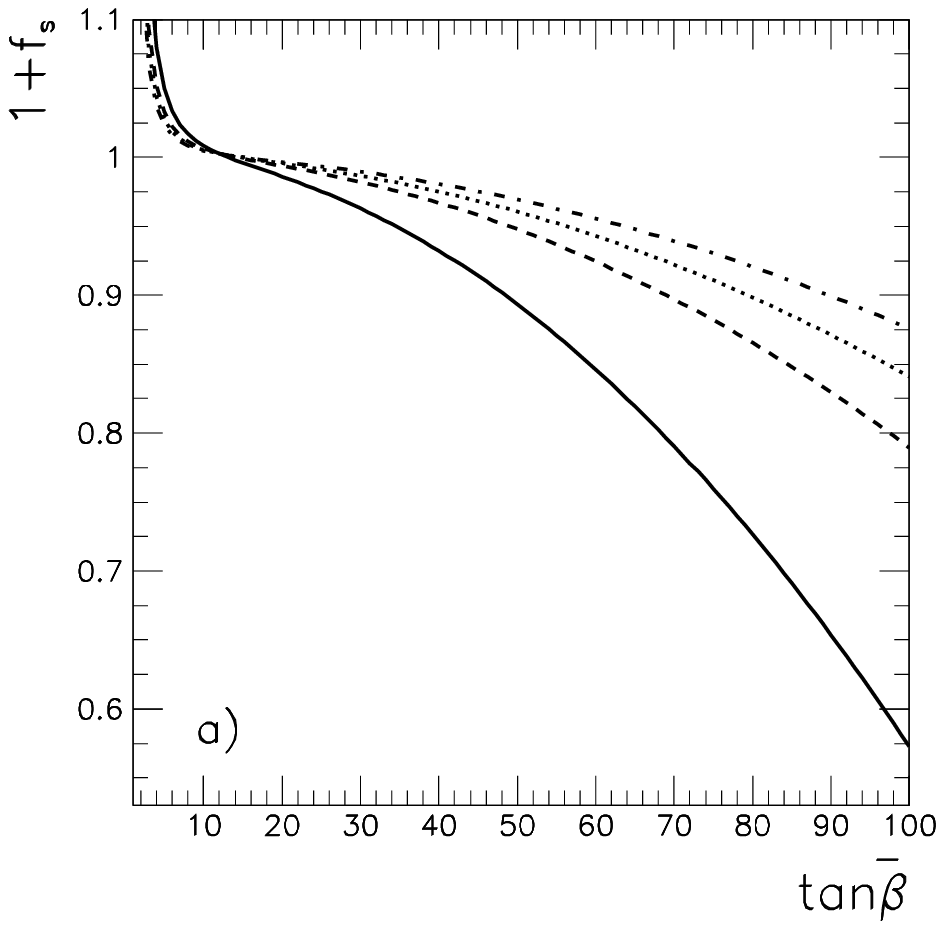,width=\linewidth}}&
\mbox{\epsfig{file=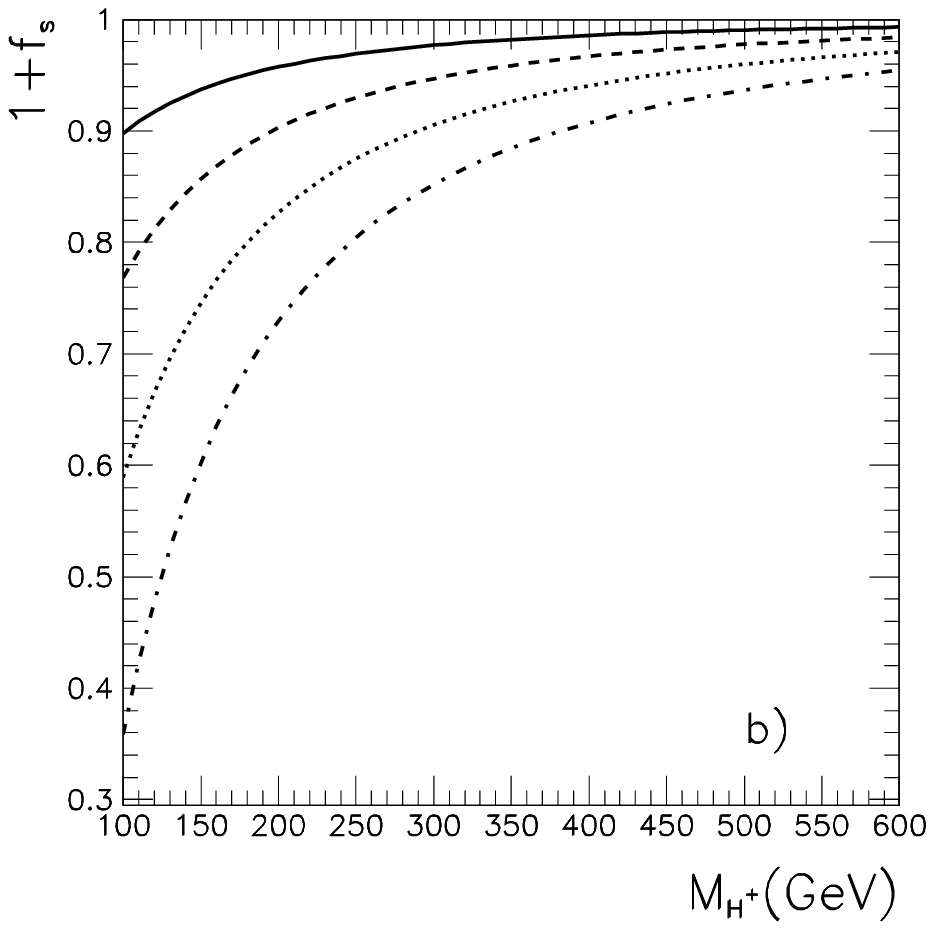,width=\linewidth}}\\
\end{tabular}
\caption{$1+f_s$ in the 2HDM(II): a) as a function of $\tan\bar\beta$ 
for $M_{H^+}=$ (from below) $150$, $250$, $300$ and $350$ GeV 
and b) as a function of $M_{H^+}$ for $\tan\bar\beta=$ (from above) $40$,
$60$, $80$ and $100$.
\label{fig:bcrs9}}
\end{center}
\end{figure}

Unfortunately, recent refinements in the computation of the
$b\rightarrow s\gamma$ rate \cite{GAMI} together with the new CLEO
experimental result for this process \cite{CLEOMOR} $BR(B\rightarrow
X_s\gamma)=(3.03\pm0.40\pm0.26)\times10^{-4}$ set the bound
$M_{H^+}\simgt380$ GeV \cite{GAMOR,GAMI}. This means that in the
2HDM(II) for the still allowed range of charged Higgs boson masses the
decrease of $1+f_s$ can be very small. Consequently, the SM analysis
of the unitarity triangle based on $\varepsilon$, $\Delta M_d$ and
$\Delta M_s$ is practically unchanged in the 2HDM(II) for large
$\tan\bar\beta\simlt50$.

However, the bound on $M_{H^+}$ from the $b\rightarrow s\gamma$ rate does 
not apply in the MSSM which we consider in the next subsection.

\subsection{MSSM with large $\tan\bar\beta$}

As a second realistic GMFV model we consider the MSSM. At the one loop level 
the contributions to the Wilson coefficients of the $|\Delta F|=2$ operators 
(\ref{eqn:ops}) in the MSSM are 
given\footnote{In accordance with the general 
framework of this paper we assume here that the CKM matrix is the only source
of flavour and CP violation.} by chargino-top squark box diagrams and by
box diagrams with the charged Higgs boson. Since the Higgs sector of the MSSM
is (at the tree level) a special case of the general 2HDM(II) considered
in subsection 4.1, the latter contribution is described by the formulae
(\ref{eqn:WHcontr}), (\ref{eqn:HHcontr}). It is well known 
\cite{MIPORO,BUGAGOJASI2,GAGI} 
that for $\tan\bar\beta$ not too big, the MSSM
is of the MFV type and both, the chargino-stop and the Higgs sectors give
{\sl positive} contributions to $1+f_s\approx1+f_d\approx1+f_\varepsilon$
which are the bigger the lighter are particles in the loops and the smaller 
is the value of $\tan\bar\beta$. In this section we want to consider the
MSSM with large values of $\tan\bar\beta$ which are favoured both by the 
LEP limit on the mass of the lighter neutral Higgs boson and by the recent 
measurement of the anomalous magnetic moment of the muon \cite{BR}.
As the treatment of the full MSSM contribution is complicated, we 
will for simplicity consider here only the limit of heavy sparticles and 
will concentrate only on the most spectacular effects.
Complete analysis of the MSSM will be presented elsewhere \cite{BUCHROSL}.

In the limit of heavy sparticles (which is practically realized
already for $M_{\rm sparticles}\simgt500$ GeV) the one loop diagrams
involving charginos and stops are negligible. It is however known that
for large $\tan\bar\beta$ even if sparticles are heavy they can still
compensate the $H^\pm$ contribution to the $b\rightarrow s\gamma$
amplitude allowing for the existence of a light, $\sim{\cal O}(150$
GeV), charged Higgs boson \cite{DEGAGI,CAGANIWA}.  From
fig.~\ref{fig:bcrs9} it follows therefore that, even for\footnote{In
the MSSM one usually constrains $\tan\bar\beta$ to be less than
$50-55$ by requiring perturbativity of the Yukawa couplings up to the
GUT scale $\sim10^{16}$ GeV.}  $\tan\bar\beta\simlt50$ and already at
the one loop level the contribution of the MSSM Higgs sector to the
$C_2^{\rm LR}$ Wilson coefficient can be non-negligible.

At the two loop level one has to take into account not only the ${\cal
O}(\alpha_s)$ corrections to the Wilson coefficients (which in the
MSSM arise from exchanges of gluons as well as gluinos) but also the
dominant two loop electroweak corrections (proportional to large top
and, in the case of large $\tan\bar\beta$, bottom Yukawa couplings).
The gluonic ${\cal O}(\alpha_s)$ corrections to the charged Higgs box
diagrams are expected to be of the same order of magnitude as the
gluonic correction to the SM $t-W^\pm$ box diagrams, i.e. 
moderate~\cite{URKRJESO}.  Also most of the two loop diagrams
involving sparticles will give contributions suppressed by the inverse
of the large sparticle masses.  The one loop effects of the MSSM Higgs
sector are however enhanced by an important class of two loop
corrections involving sparticles.  In the limit of heavy sparticles
these corrections can be most easily identified in the effective
Lagrangian approach \cite{CHPO}.  While direct contribution of heavy
sparticles to the Wilson coefficients can be neglected, it is well
known that for large $\tan\bar\beta$ such sparticles do not decouple
entirely \cite{HARASA}. Integrating them out modifies the original
couplings of the two Higgs doublets to the known fermions leading to
the following quark Yukawa interactions:
\begin{eqnarray}
&&{\rm L}_{\rm Yuk} = -\epsilon_{ij}H^{(d)}_i d_I^c Y_d^{IA} q_{Aj}
                    -H^{(u)\ast}_i d_I^c (\Delta_uY_d)^{IA} q_{Ai}\phantom{aa}
\nonumber\\
&&\phantom{aaaa} -\epsilon_{ij}H^{(u)}_i u_A^c Y_u^{AI} q_{Ij}
                 -H^{(d)\ast}_i u_A^c(\Delta_dY_u)^{AI} q_{Ii} + {\rm H.c.}
\label{eqn:Leff}
\end{eqnarray} 
where $H^{(d)}_i$ and $H^{(u)}_i$ are the two Higgs doublets giving at
the tree level masses to the down- and up-type quarks, respectively
and $q_i$, $d^c$ and $u^c$ are the fields of the left handed fermions
(for simplicity we use here the Weyl spinors). Dominant corrections
$(\Delta_uY_d)^{IA}$ and $(\Delta_dY_u)^{AI}$ are finite and
calculable in terms of the sparticle parameters. In the effective
Lagrangian approach they can be obtained from diagrams shown in
fig.~\ref{fig:bcrs10} in which $Q$, $U^c$ and $D^c$ are the scalar
superpartners of the SM fermions $q_i$, $d^c$ and $u^c$, $\tilde g$ is
the gluino and $\tilde H^{(d)}_i$ and $\tilde H^{(u)}_i$ are the
fermionic superpartners of the two Higgs doublets.

\begin{figure}[htbp] 
\begin{center}
\begin{picture}(400,300)(0,0)
\ArrowLine(90,180)(120,180)
\ArrowLine(90,180)(60,180)
\ArrowLine(10,180)(60,180)
\ArrowLine(170,180)(120,180)
\Line(87,183)(93,177)
\Line(87,177)(93,183)
\Text(20,175)[t]{$q_A$}
\Text(160,175)[t]{$d^c_I$}
\Text(90,175)[t]{$\tilde{g}$}
\Text(73,200)[t]{$Q_B$}
\Text(107,200)[t]{$D^c_J$}
\Text(70,270)[t]{$H^{(u)}$}
\DashArrowLine(50,180)(90,240){4}
\DashArrowLine(130,180)(90,240){4}
\DashArrowLine(90,240)(90,275){4}
\Text(90,155)[t]{$a)$}
\ArrowLine(290,180)(320,180)
\ArrowLine(290,180)(260,180)
\ArrowLine(210,180)(260,180)
\ArrowLine(370,180)(320,180)
\Line(287,183)(293,177)
\Line(287,177)(293,183)
\Text(220,175)[t]{$q_A$}
\Text(360,175)[t]{$d^c_I$}
\Text(275,175)[t]{$\tilde H^{(u)}$}
\Text(310,175)[t]{$\tilde H^{(d)}$}
\Text(273,200)[t]{$U^c_B$}
\Text(310,200)[t]{$Q_J$}
\Text(270,270)[t]{$H^{(u)}$}
\DashArrowLine(290,240)(250,180){4}
\DashArrowLine(290,240)(330,180){4}
\DashArrowLine(290,240)(290,275){4}
\Text(290,155)[t]{$b)$}
\ArrowLine(90,30)(120,30)
\ArrowLine(90,30)(60,30)
\ArrowLine(10,30)(60,30)
\ArrowLine(170,30)(120,30)
\Line(87,33)(93,27)
\Line(87,27)(93,33)
\Text(20,25)[t]{$q_I$}
\Text(160,25)[t]{$u^c_A$}
\Text(90,25)[t]{$\tilde{g}$}
\Text(73,50)[t]{$Q_J$}
\Text(107,50)[t]{$U^c_B$}
\Text(70,120)[t]{$H^{(d)}$}
\DashArrowLine(50,30)(90,90){4}
\DashArrowLine(130,30)(90,90){4}
\DashArrowLine(90,90)(90,125){4}
\Text(90,5)[t]{$c)$}
\ArrowLine(290,30)(260,30)
\ArrowLine(290,30)(320,30)
\ArrowLine(210,30)(260,30)
\ArrowLine(370,30)(320,30)
\Line(287,33)(293,27)
\Line(287,27)(293,33)
\Text(220,25)[t]{$q_I$}
\Text(360,25)[t]{$u^c_A$}
\Text(275,25)[t]{$\tilde H^{(d)}$}
\Text(310,25)[t]{$\tilde H^{(u)}$}
\Text(273,50)[t]{$D^c_B$}
\Text(310,50)[t]{$Q_J$}
\Text(270,120)[t]{$H^{(d)}$}
\DashArrowLine(290,90)(250,30){4}
\DashArrowLine(290,90)(330,30){4}
\DashArrowLine(290,90)(290,125){4}
\Text(290,5)[t]{$d)$}
\end{picture}
\end{center}
\caption{Diagrams giving rise to dominant, $\tan\bar\beta$ enhanced, 
corrections $\Delta_uY_d$ (diagrams a and b) and $\Delta_dY_u$ 
(diagrams c and d).}
\label{fig:bcrs10}
\end{figure}
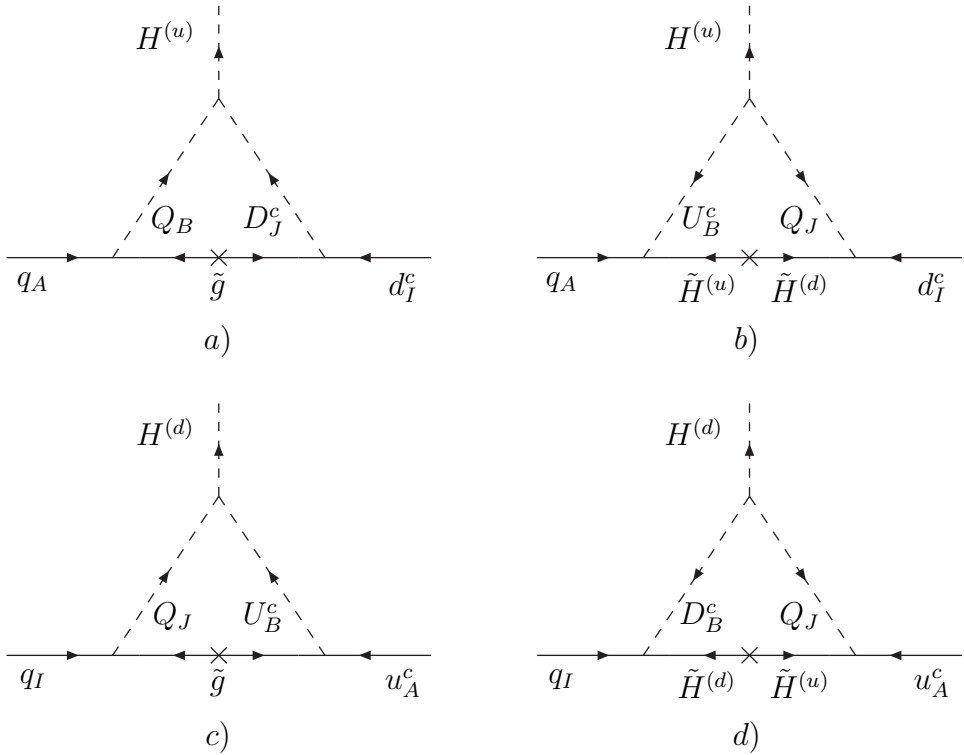

The effects of the corrections $\Delta Y$ are threefold. Firstly, they
modify the 2HDM(II) relations between the masses $m_{d_I}(\mu_t)$ 
and the eigenvalues of the Yukawa matrices $Y_d^{IA}$:
\begin{eqnarray}
Y_d^I = {e\over\sqrt2s_W}{m_{d_I}\over M_W}\sqrt{1+\tan^2\bar\beta}
\rightarrow{e\over\sqrt2s_W}{m_{d_I}\over M_W}
{\sqrt{1+\tan^2\bar\beta}\over1+\epsilon_d(I)\tan\bar\beta}
\label{eqn:Ymrel}
\end{eqnarray} 
where $\epsilon_d(I)$ can be found in ref. \cite{DEGAGI}. Secondly, they 
induce additional, $\propto\tan\bar\beta$, terms in the couplings 
(\ref{eqn:Lcouplings}), (\ref{eqn:Rcouplings}) \cite{CIDEGAGI,DEGAGI}. 
Both these corrections should be taken into account in vertices of the 
charged Higgs boson box diagrams and constitute, therefore, the two loop
corrections to the Wilson coefficients of the $|\Delta F|=2$ operators.
It turns out, however, that the most 
important (for non-negligible mixing of the top squarks) effect of the 
corrections $\Delta Y$ is the generation of the flavour non-diagonal, 
$\tan\bar\beta$ enhanced couplings of the neutral Higgs bosons to down-type 
quarks \cite{HAPOTO,BAKO}. In the effective Lagrangian approach these 
couplings originate from the diagram \ref{fig:bcrs10}b. Details of their
calculation have been presented in ref.~\cite{CHSL}. They can be also computed
diagramatically as in \cite{CHSL,BOEWKRUR} what allows to take 
fully into account the complicated composition of charginos. Additional 
contributions to the Wilson coefficients $C^{\rm SLL}_1$, $C^{\rm SRR}_1$ 
and $C^{\rm LR}_2$ are then generated by the double penguin 
diagrams shown in fig.~\ref{fig:bcrs11} in which the neutral Higgs bosons
are exchanged at the tree level between two effective flavour
changing vertices generated at one loop. 

The single neutral Higgs penguin 
diagram for the $(d_I)_{\rm L(R)}\rightarrow (d_J)_{\rm R(L)}$ transition 
(where L and R refer to the quark chiralities) grows as $\tan^2\bar\beta$ 
and is proportional to $m_{d_J}$ ($m_{d_I}$) \cite{CHSL}. Consequently the 
double penguin diagrams in fig.~\ref{fig:bcrs11} grow like
$\tan^4\bar\beta$ times $m_{d_J}^2$, $m_{d_I}^2$ and $m_{d_J}m_{d_I}$, 
respectively. This mismatch of powers in $\tan\bar\beta$ and powers of
light quark masses in Higgs penguin diagrams should be contrasted with box
diagrams, where each $\tan\bar\beta$ is accompanied by a light quark mass
as seen in (\ref{eqn:WHorder})-(\ref{eqn:BBorder}). In this manner the 
two-loop electroweak double
Higgs penguins can potentially compete with the one loop electroweak box
diagrams. In the standard diagramatic approach to the calculation of the 
flavour changing $d_Jd_IH^0(h^0,A^0)$ vertex the mismatch of powers in 
question can be understood simplest as follows. The diagrams for the genuine 
1-PI vertex corrections (with physical squarks and charginos in the loops) 
contribute only terms proportional to one power of $\tan\bar\beta$. 
They are however accompanied by two tree level
flavour conserving $d_J d_J$-Higgs and $d_I d_I$-Higgs  vertices with 
flavour violating self-energies (involving squarks and charginos) 
on the external quarks lines. As the external momenta can be neglected, 
the internal fermion propagators of $d_J$ and $d_I$ cancel respectively 
the $m_{d_J}$ and $m_{d_I}$ factors present in the
quark-quark-Higgs vertices.

For the transitions $d_I\bar d_J\rightarrow d_J\bar d_I$ the dominant terms 
obtained from the double penguin contributions are
\begin{eqnarray}
&&\delta^{(0)}C^{\rm SLL}_1=-{\alpha_{EM}\over4\pi s^2_W}{m^4_t\over M^4_W}
m_{d_J}^2 X_{tC}^2 \tan^4\bar\beta ~{\cal F}_-\phantom{aa}\nonumber\\ 
&&\delta^{(0)}C^{\rm SRR}_1=-{\alpha_{EM}\over4\pi s^2_W}{m^4_t\over M^4_W}
m_{d_I}^2 X_{tC}^2 \tan^4\bar\beta ~{\cal F}_-\phantom{aa}\\ 
&&\delta^{(0)}C^{\rm LR}_2=-{\alpha_{EM}\over2\pi s^2_W}{m^4_t\over M^4_W}
m_{d_J}m_{d_I} X_{tC}^2 \tan^4\bar\beta ~{\cal F}_+~.\phantom{aa}\nonumber
\label{eqn:babucor}
\end{eqnarray}
$X_{tC}$ is given by
\begin{eqnarray}
X_{tC} = \sum_{j=1}^2Z_+^{2j}Z_-^{2j}{A_t\over m_{C_j}}
H_2(x^{t/C_j}_1,x^{t/C_j}_2),
\end{eqnarray}
where $x^{t/C_j}_i=M^2_{\tilde t_i}/m^2_{C_j}$, $i=1,2$, $j=1,2$
are the ratios of the stop and chargino masses squared, the matrices
$Z_+$ and $Z_-$ are defined in ref.~\cite{ROS} and the 
function $H_2(x,y)$ is defined in the Appendix. The factor
\begin{eqnarray}
{\cal F}_\mp\equiv\left[{\cos^2\bar\alpha\over M^2_H} 
+ {\sin^2\bar\alpha\over M^2_h} \mp {\sin^2\bar\beta\over M^2_A}\right]
\end{eqnarray}
depends on the masses of the CP-even neutral Higgs bosons $h^0$ and $H^0$, 
the mass of the CP-odd Higgs boson $A^0$ (in the MSSM 
$M^2_{H^+}=M^2_A+M^2_W$) and the mixing angles $\bar\alpha$ and $\bar\beta$.
For $\tan\bar\beta\gg1$ and $M_A\simgt130$ GeV, $\cos^2\bar\alpha\approx1$,
$\sin^2\bar\alpha\approx0$ and $M_H\approx M_A$. 

\begin{figure}[htbp]
\begin{center}
\begin{picture}(340,110)(0,0)
\ArrowLine(10,10)(50,30)
\ArrowLine(50,30)(90,10)
\Vertex(50,30){7}
\ArrowLine(50,80)(10,100)
\ArrowLine(90,100)(50,80)
\Vertex(50,80){7}
\DashLine(50,30)(50,80){3}
\Text(78,55)[]{$h^0$,$H^0$,$A^0$}
\Text(25,05)[]{$(d_L)_I$}
\Text(75,05)[]{$(d_R)_J$}
\Text(75,105)[]{$(d_L)_I$}
\Text(25,105)[]{$(d_R)_J$}
\ArrowLine(130,10)(170,30)
\ArrowLine(170,30)(210,10)
\Vertex(170,30){7}
\ArrowLine(170,80)(130,100)
\ArrowLine(210,100)(170,80)
\Vertex(170,80){7}
\DashLine(170,30)(170,80){3}
\Text(198,55)[]{$h^0$,$H^0$,$A^0$}
\Text(145,05)[]{$(d_R)_I$}
\Text(195,05)[]{$(d_L)_J$}
\Text(195,105)[]{$(d_R)_I$}
\Text(145,105)[]{$(d_L)_J$}
\ArrowLine(250,10)(290,30)
\ArrowLine(290,30)(330,10)
\Vertex(290,30){7}
\ArrowLine(290,80)(250,100)
\ArrowLine(330,100)(290,80)
\Vertex(290,80){7}
\DashLine(290,30)(290,80){3}
\Text(318,55)[]{$h^0$,$H^0$,$A^0$}
\Text(265,05)[]{$(d_L)_I$}
\Text(315,05)[]{$(d_R)_J$}
\Text(315,105)[]{$(d_R)_I$}
\Text(265,105)[]{$(d_L)_J$}
\end{picture}
\end{center}
\caption{Double penguin diagram contributing to: a) $C_1^{\rm SLL}$,
  b) $C_1^{\rm SRR}$ and c) $C_2^{\rm LR}$ Wilson coefficients in the
  MSSM with large $\tan\bar\beta$.}
\label{fig:bcrs11}
\end{figure}
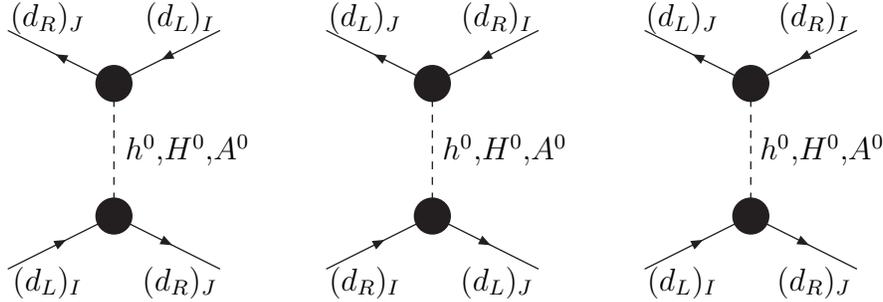

\begin{figure}[htbp]
\begin{center}
\epsfig{file=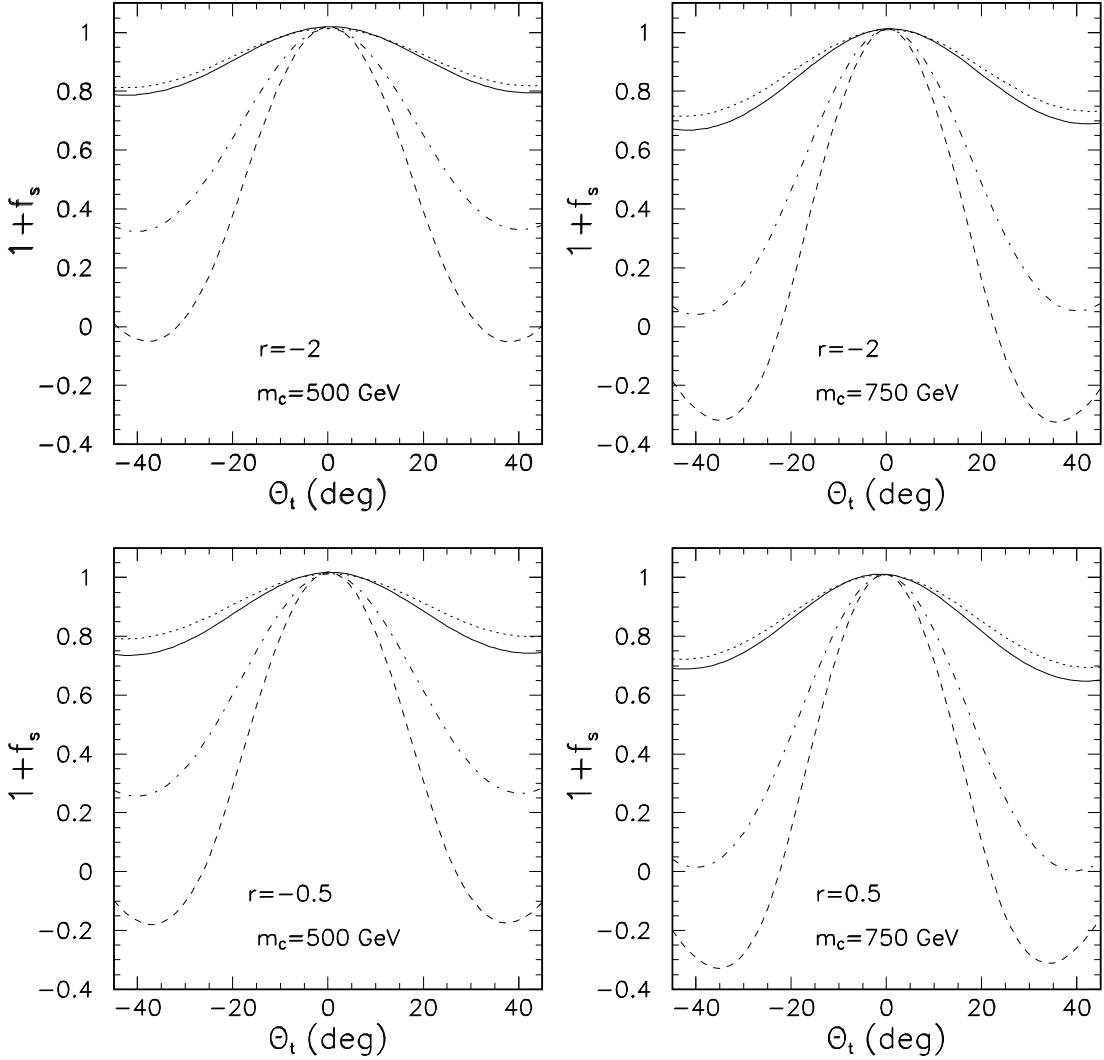,width=\linewidth}
\end{center}
\caption{\protect $1+f_s$ in the MSSM as a function of the mixing angle
  of the top squarks for different lighter chargino masses and
  compositions ($r\equiv M_2/\mu$). Solid, dashed, dotted and
  dot-dashed lines correspond to stop masses (in GeV) (500,650),
  (500,850), (700,850) and (700,1000), respectively.}
\label{fig:bcrs12}
\end{figure}

\begin{figure}[htbp]
\begin{center}
\epsfig{file=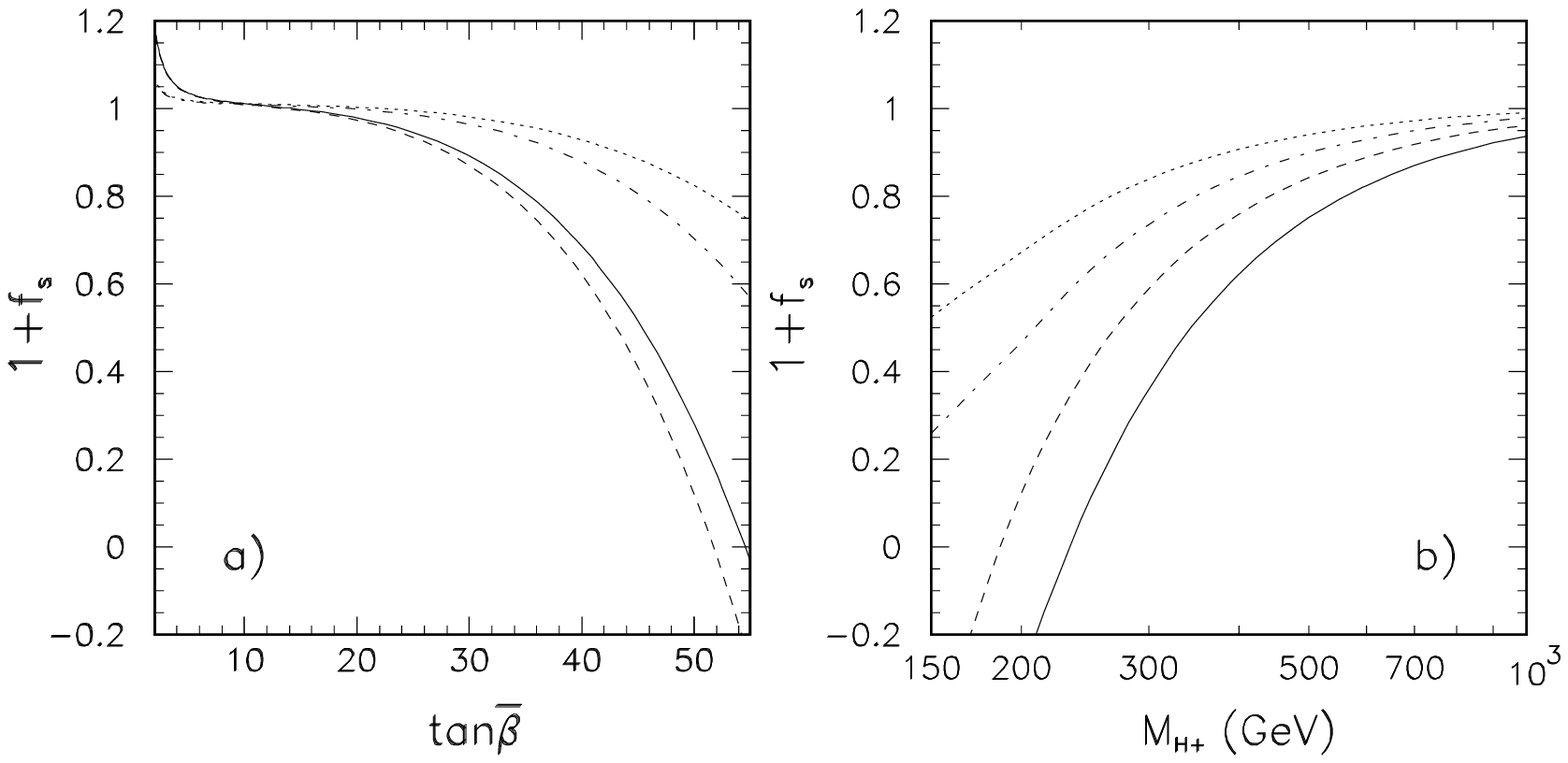,width=\linewidth}
\end{center}
\caption{\protect $1+f_s$ in the MSSM for lighter chargino mass
  750 GeV, $r\equiv M_2/\mu=-0.5$ and stop masses (in GeV) (500,850),
  (700,1000), (500,850) and (600,1100) (solid, dashed, dotted and
  dot-dashed lines, respectively) as a function of a) $\tan\bar\beta$ and
  b) $M_{H^+}$.  In panel a) solid and dashed (dotted and dot-dashed)
  lines correspond to $M_{H^+}=200$ $(600)$ GeV, and in panel a) solid
  and dashed (dotted and dot-dashed) lines correspond to
  $\tan\bar\beta=50$ $(35)$.}
\label{fig:bcrs13}
\end{figure}

As has been observed in \cite{BAKO}, $\delta^{(0)}C^{\rm SLL}_1$
and $\delta^{(0)}C^{\rm SRR}_1$ depend on ${\cal F}_-$ which for
$\tan\bar\beta\gg1$ is close to zero and strongly suppresses
these corrections. However the correction $\delta^{(0)}C^{\rm LR}_2$
is proportional to ${\cal F}_+$ which is not suppressed in this limit.
Approximating for simplicity the dimensionless factor $X_{tC}$ by unity,
it is easy to see that in the case of the \BBs mixing 
this correction, although proportional to small strange quark mass
($m_s(\mu_t)\approx61$ MeV), can be for $\tan\bar\beta\sim50$ and 
$M_{H^+}\sim200$ GeV as large as $\delta^{(0)}C^{\rm LR}_2\sim2.5$ i.e. of 
the same order of magnitude as the SM contribution to $C^{\rm VLL}_1$.
Consequently there can be a significant, growing as $\tan^4\bar\beta$,
contribution to the $C_2^{\rm LR}$ 
Wilson coefficient which is further enhanced (relative to the 
$C_1^{\rm VLL}$ coefficient) by QCD effects (see sec. 3). 

An important feature of the double penguin contribution is its fixed
negative sign (because it is proportional $X_{tC}^2$) i.e. the same as
the sign of the dominant effects of the charged Higgs box diagrams at
large $\tan\bar\beta$.  Therefore the double penguin contribution
interferes destructively with the SM contribution and leads to
$1+f_s<1$. Another interesting feature is its strong dependence on the
left-right mixing of the top squarks which is clearly visible in
fig.~\ref{fig:bcrs12} where we show $1+f_s$ as a function of the stop
mixing angle $\theta_t$ for different chargino masses and compositions
and different choices of the stop masses. For the same value of the
mixing angle $\theta_t$, larger effects are obtained for bigger stop
mass splitting because in this case the parameter $|A_t|$ has to be
larger.  It should be also stressed that this contribution does not
vanish when the mass scale of the sparticles is increased (i.e. when
all mass parameters are scaled uniformly). Thus, large effects
decreasing $1+f_s$ below unity can be present in the MSSM also for the
heavy sparticles provided the mass scale of the MSSM Higgs sector
remains low and $\tan\bar\beta$ is large. This is illustrated in
figs.~\ref{fig:bcrs13}a and b where we show $1+f_s$ as a function of
$\tan\bar\beta$ (panel a) and $M_{H^+}$ (panel b). Positive
contribution to $1+f_s$ seen in fig.~\ref{fig:bcrs13}a for
$\tan\bar\beta<2.5$ and $M_{H^+}=200$ GeV is due to the ordinary
charged Higgs box diagrams which contribute to the universal part of
$f_s$, $f_d$ and $f_\varepsilon$ through the Wilson coefficient of the
standard $Q_1^{\rm VLL}$ operator.  For lighter $H^+$ and light
charginos $1+f_s$ can reach values $\sim2$ \cite{BUGAGOJASI2}. As
follows from (\ref{eqn:1pfsbound}) such high values of $1+f_s$ could be
soon excluded by the measurement of $\Delta M_s$.

The implications of these results are as follows. A large portion of the MSSM 
parameter space considered here leads to $1+f_s$ values violating already 
at present the bound (\ref{eqn:1pfsbound}) and is, hence, excluded. It is 
particularly interesting that part of this parameter space is not yet excluded 
by the recent results for the $b\rightarrow s\gamma$ rate. This is because 
the theoretical prediction for the latter depends, besides stop and chargino 
sector parameters (on which the double penguin contribution does depend), 
also on masses of sfermions from the first two generations, the gluino mass 
etc. This can have important consequences for other processes involving 
the $B^0$ mesons. For instance, orders of magnitude enhancement of the 
$B^0_{s,d}\rightarrow\mu^+\mu^-$ branching ratio found in the MSSM for very 
large mixing of stops \cite{BAKO,CHSL,BOEWKRUR,HULIYAZH} is ruled out, at 
least for sparticles so heavy that their direct contribution to the 
$\bar B^0_s$-$B^0_s$  mixing cannot significantly reduce the negative 
double penguin contribution to $1+f_s$. Thus, finding
$BR(B^0_{s,d}\rightarrow\mu^+\mu^-)$ close to the present experimental
bound and no light stops and charginos would strongly indicate a non-minimal 
flavour violation in supersymmetry \cite{CHSL}.

Furthermore, for the MSSM parameters, for which the bound
(\ref{eqn:1pfsbound}) is respected,
$1+f_d\approx1+f_\varepsilon\approx1$ and in this sector the MSSM in
the limit $M_W\simlt M_{H^+}\ll M_{\rm sparticle}$ mimics the SM. In
particular, it is still consistent with the present experimental data
for $\varepsilon$, $\Delta M_d$, $\Delta M_s$ and $a_{\psi K_{\rm
S}}$. Measuring $\Delta M_s$ larger than the present lower bound
$\Delta M_s>15/ps$ will further limit allowed combinations of stop
mixings, their mass splittings and chargino parameters. Only very
large values of $\Delta M_s$, requiring $1+f_s\simgt1$, would rule out
the supersymmetric scenario considered here entirely; the SM would be
then ruled out too.

Obviously, finding the asymmetry $a_{\psi K_{\rm S}}$ below its SM
value would also rule out this scenario entirely because for
$1+f_d\approx1+f_\varepsilon\approx1$ the unitarity of the CKM matrix
requires $\sin2\beta\simgt0.5$, i.e. bigger than the bound
(\ref{bound}) valid in MVF models which admit
$1+f_d\approx1+f_\varepsilon\neq1$. If $a_{\psi K_{\rm S}}$ is found
around $0.6$ then the combination of constraints from $R_b$,
$\varepsilon$ and $\sin2\beta$ similar to the ones shown in
fig.~\ref{fig:bcrs8} can put slightly stronger limits on $1+f_s\approx
R_{sd}$ than the bound (\ref{eqn:1pfsbound}) alone but the usefulness
of this limits will depend crucially on the measured value of $\Delta
M_s$.

Finally, if experimentally $a_{\psi K_{\rm S}}\approx0.7$ and $\Delta
M_s$ combined with improved lattice results for $\sqrt{\hat
B_{B_s}}F_{B_s}$ and $\xi$ allow for $R_{sd}\approx1+f_s\sim0.65-0.8$,
this scenario can lead to angle $\gamma$ moderately smaller than the
one predicted in the SM.

\section{Conclusions}
\setcounter{equation}{0}

In this paper we have investigated the role of new dimension six
four-fermion $|\Delta F|=2$ operators in models with minimal flavour
violation. Short distance contributions to the mass differences
$\Delta M_s$, $\Delta M_d$ and to the CP violation parameter
$\varepsilon$ are parameterized by three real functions $F^s_{tt}$,
$F^d_{tt}$ and $F^\varepsilon_{tt}$, respectively. General formulae
for $F^i_{tt}$ in terms of the Wilson coefficients evaluated at the
scale $\mu=\mu_{\rm NP}$, the relevant QCD renormalization group
factors and the non-perturbative $B_i$-factors have been presented.

We have proposed a few simple strategies involving the ratio $\Delta
M_s/\Delta M_d$, $\sin2\beta$ and the angle $\gamma$ that allow to
search for the effects of the new operators. We have also found model
independent bounds on the functions $1+f_i=F^i_{tt}/S_0(x_t)$ that
should be considerably improved once $\Delta M_s/\Delta M_d$, $\sin
2\beta$ and the angle $\gamma$ are precisely measured and our
knowledge about non-perturbative parameters and the CKM elements
$|V_{ub}|$ and $\vcb$ is improved.  Our findings can be summarized as
follows:

\begin{itemize}
\item The present experimental and theoretical uncertainties allow for
  sizable contributions of new operators to $\Delta M_{s,d}$ and
  $\varepsilon$.
\item As the unitarity of the CKM matrix implies $|V_{ts}|\approx
  \vcb$ independently of new physics contributions, the function
  $1+f_s$ can be determined from the experimental value of $\Delta
  M_s$ subject to the uncertainties in $\vcb$, $m_t$ and in particular
  $\sqrt{\hat B_{B_s}}F_{B_s}$. For instance for $\Delta M_s=18/ps$ we
  find $0.63\le 1+f_s\le 1.55$. The decrease of the theoretical error
  in $\sqrt{\hat B_{B_s}}F_{B_s}$ accompanied by a precise measurement 
  of $\Delta M_s$ should tell us whether $1+f_s >1$ or $1+f_s<1$ thereby
  excluding certain scenarios and putting important constraints on the
  parameter space of the surviving models.
\item We find that values of $R_{sd}=(1+f_s)/(1+f_d)$ substantially
  different from unity would allow $\sin2\beta$ to be lower than in
  the MFV models and in particular in the SM. Simultaneously
  $R_{d\varepsilon}=(1+f_d)/(1+f_\varepsilon) < 1$ would be favoured.
\item Whether $R_{sd}>1$ or $R_{sd}<1$ is favoured by the data can be
  decided by the measurement of the angle $\gamma$ with
  $\gamma>90^\circ$ and $\gamma<90^\circ$ corresponding for $\Delta
  M_s=15/ps$ to $R_{sd}>1.2$ and $R_{sd}<1.2$, respectively. For a
  given $R_{sd}$ and $\sin2\beta$, the predicted angle $\gamma$
  decreases with increasing $\Delta M_s$. For $R_{sd}>1.5$ and $\Delta
  M_s/\Delta M_d\le 40$ values of the angle $\gamma$ greater than
  $90^\circ$ are possible.
\item We have determined the presently allowed ranges in the
  $(1+f_d,1+f_\varepsilon)$ and $(R_{sd},1+f_\varepsilon)$ planes.  
  An analysis of a hypothetical measurements of $\Delta M_s$ and 
  $a_{\psi K_s}$ that allow the determination of the unitarity triangle
  illustrated various possibilities further.
\end{itemize}

As an example we have analyzed the role of new operators in the MSSM
with large $\tan\bar\beta=v_2/v_1$ in the limit of heavy sparticles,
investigating in particular the impact of the extended Higgs sector on
the unitarity triangle. Here our findings are as follows:

\begin{itemize}
\item The largest effects of new contributions for large
  $\tan\bar\beta$ are seen in $\Delta M_s$. The corresponding
  contributions to $\Delta M_d$ and $\varepsilon$ are strongly
  suppressed either by inverse powers of $\tan\bar\beta$ or by the
  smallness of $d$-quark mass.
\item The dominant contributions to $\Delta M_s$ for large
  $\tan\bar\beta$ come from the operator $Q_2^{\rm LR}=(\bar
  b(1-\gamma_5)s)(\bar b(1+\gamma_5)s)$.  They originate from double
  penguin diagrams involving neutral Higgs particles and, to a lesser
  extent, in the box diagrams with charged Higgs exchanges. The
  dominant double penguin diagrams arise through the generation of
  flavour non-diagonal $\tan\bar\beta$ enhanced couplings of neutral
  Higgs bosons to the down-type quarks and depend strongly on the
  mixing of the top squarks and their mass splitting.
\item The contribution of double penguins grows like $\tan^4\bar\beta$
  and interferes destructively with the SM contribution, suppressing
  considerably $1+f_s$ below unity.
\item All these findings have the following phenomenological
  consequences. The MSSM with large $\tan\bar\beta$, substantial stop
  mixing and large stop mass splitting realizes the $R_{sd}<1$
  scenario with $\gamma<90^\circ$ and generally smaller than in the SM
  and MFV models. As $R_{d\varepsilon}=1$ and $1+f_\varepsilon=1$,
  the lower bound $\sin 2\beta>0.50$ valid in the SM remains unchanged.
  Consequently if $a_{\psi K_S}$ is found below $0.50$, this scenario of
  the MSSM will be excluded (together with the SM) while other MSSM
  scenarios with lighter sparticles and lower $\tan\bar\beta$,
  belonging to the MFV class, may still be consistent with the data.
  As seen in fig.~\ref{fig:bcrs4}, for higher values of $a_{\psi K_S}$
  the MSSM scenario considered here is a vital possibility with the
  angle $\gamma$ smaller than in the SM, although values of $R_{sd}$
  as low as 0.6 appear rather improbable.
\item The constraint (\ref{eqn:1pfsbound}), which basically limits the 
 magnitude of the stop mixing parameter $A_t$, has also important 
 consequences for other processes involving the $B^0$ mesons. For example, 
 in the scenario considered in this paper, it severely limits possible 
 enhancement of the $B^0_{s,d}\rightarrow\mu^+\mu^-$ decay rate.
\end{itemize}
Detailed analysis of $\Delta M_{d,s}$ and $\varepsilon$ in the MSSM at large 
$\tan\bar\beta$, including also scenarios with light sparticles 
will be presented soon \cite{BUCHROSL}.

It will be exciting to watch future developments in the experimental
values of $\Delta M_s$, $a_{\psi K_S}$ and the angle $\gamma$ that 
will either choose one of the possibilities considered in this paper
or constrain the parameters of GMFV models.

\vskip 1cm

\noindent {\bf Acknowledgments}
\vskip 0.3cm
\noindent 
A.J.B. and J.R. would like to thank Sebastian J\"ager, Frank Kr\"uger,
J\"org Urban, Enrico Lunghi and Oscar Vives for useful discussions.
The work of A.J.B. and J.R. was supported in part by the
German Bundesministerium f\"ur Bildung and Forschung under the
contract 05HT1WOA3. The work of P.H.Ch. was supported partially by the
Polish State Committee for Scientific Research grant 5 P03B 119 20 for
2001-2002 and by the EC Contract HPRN-CT-2000-00148 for years
2000-2004.  J.R. was also partially supported by the Polish State
Committee for Scientific Research grant 2 P03B 060 18 for years
2000-2001.

\renewcommand{\thesection}{Appendix~\Alph{section}}
\renewcommand{\theequation}{\Alph{section}.\arabic{equation}}

\setcounter{section}{0}
\setcounter{equation}{0}

\section{}
\label{sec:suscon}
The four point loop functions with zero external
momenta  are defined as follows:
\begin{eqnarray}
D_0(a,b,c,d)&=&\int{d^4k\over\pi^2}{i\over[a][b][c][d]}
={b\over (b-a)(b-c)(b-d)}\log{b\over a}\nonumber\\
&+&{c\over (c-a)(c-b)(c-d)}\log{c\over a}
+{d\over (d-a)(d-b)(d-c)}\log{d\over a}\\
D_2(a,b,c,d)&=&\int{d^4k\over\pi^2}{ik^2\over[a][b][c][d]}
={b^2\over (b-a)(b-c)(b-d)}\log{b\over a}\nonumber\\
&+&{c^2\over (c-a)(c-b)(c-d)}\log{c\over a}
+{d^2\over (d-a)(d-b)(d-c)}\log{d\over a}
\label{eqn:Dfun}
\end{eqnarray}
where $[a]\equiv k^2-a$ etc. Finally,
\begin{eqnarray}
H_2(x,y)= {x\ln x\over(1-x)(x-y)}+{y\ln y\over(1-y)(y-x)}~.
\end{eqnarray}

\vfill\eject


\newpage

\begin{thebibliography}{99}
\bibitem{BUER} A.J. Buras, {\sl hep-ph}/0101336, lectures at the
  International Erice School, August, 2000.
  
\bibitem{BaBarNew} B. Aubert et al., BaBar Collaboration,
  BABAR-PUB-01-18 ({\sl hep-ex}/0107013).
  
\bibitem{BelleNew} K. Abe et al., Belle Collaboration, Belle Preprint
  2001-10 ({\sl hep-ex}/0107061).
  
\bibitem{CDF} T. Affolder et al., CDF Collaboration, {\sl Phys. Rev.}
  {\bf D61} (2000), 072005.
  
\bibitem{ALEPH} The ALEPH Collaboration, {\sl hep-ex}/0009058.
  
\bibitem{STO} LEP B-oscillation Working Group:
  http://lepbosc.web.cern.ch/LEPBOSC/combined{\_}results/.
  
\bibitem{ALLO} A. Ali and D. London, {\sl Eur. Phys. J.} {\bf C9}
  (1999), 687; {\sl Phys. Rep.} {\bf 320} (1999), 79; {\sl
    hep-ph}/0002167; {\sl Eur. Phys. J.} {\bf C18} (2001), 665.
  
\bibitem{BULAOS} A.J. Buras, M.E. Lautenbacher and G. Ostermaier, {\sl
    Phys. Rev.} {\bf D 50} (1994), 3433.
  
\bibitem{BUGAGOJASI} A.J. Buras, P. Gambino, M. Gorbahn, S. J\"ager
  and L. Silvestrini, {\sl Phys. Lett.} {\bf B500} (2001) 161.
  
\bibitem{BUBU} A.J. Buras and R. Buras, {\sl Phys. Lett.} {\bf B501}
  (2001), 223.
  
\bibitem{SM} M. Ciuchini, G. D'Agostini, E. Franco, V. Lubicz, G.
  Martinelli, F. Parodi, P.~Roudeau and A. Stocchi, {\sl
    hep-ph}/0012308; S.  Plaszczynski and M.-H. Schune, {\sl
    hep-ph}/9911280; A. H\"ocker, H.  Lacker, S. Laplace and F. Le
  Diberder, {\sl hep-ph}/0104062; S. Mele, talk Presented at 5$^{th}$
  {\sl International Symposium on Radiative Corrections (RADCOR
    2000)}, Carmel, California, September 2000, {\sl hep-ph}/0103040.
  
\bibitem{BaBar} B. Aubert et al., BaBar Collaboration, {\sl
    hep-ex}/0102030.

  
\bibitem{EYNIPE} G. Eyal, Y. Nir and G. Perez, {\sl JHEP} {\bf 0008}
  (2000), 028.
  
\bibitem{SIWO} J.P. Silva and L. Wolfenstein, {\sl Phys. Rev.} {\bf
    D63} (2001) 056001; A. Masiero, M. Piai and O. Vives, {\sl
    hep-ph}/0012096.
  
\bibitem{KANEU} A.L. Kagan and M. Neubert, {\sl Phys. Lett.} {\bf
    B492} (2000) 115.
  
\bibitem{XING} Z.Z. Xing, {\sl hep-ph}/0008018; H. Fritzsch and Z.Z. Xing,
  {\sl hep-ph}/0102295.
  
\bibitem{FLMA} R. Fleischer and Th. Mannel, {\sl hep-ph}/0101276.
  
\bibitem{WUZH} Y.-L. Wu and Y.-F. Zhou, {\sl hep-ph}/0102310.
  
\bibitem{MAVI} A. Masiero and O. Vives, {\sl hep-ph}/0104027; A. Masiero, M.
  Piai, A. Romanino and L. Silvestrini, {\sl hep-ph}/0104101.
  
\bibitem{BEPE} S. Bergmann and G. Perez, {\sl hep-ph}/0103299.

\bibitem{ALLU} A. Ali and E. Lunghi,  {\sl hep-ph}/0105200.
  
\bibitem{CIDEGAGI} M. Ciuchini, G. Degrassi, P. Gambino and G.-F.
  Giudice, {\sl Nucl. Phys.} {\bf B534} (1998),~3.
  
\bibitem{BUFL} A.J. Buras and R. Fleischer, preprint TUM-HEP-412/01
  ({\sl hep-ph}/0104238).
  
\bibitem{MIPORO} M. Misiak, S. Pokorski and J. Rosiek, in {\sl Heavy
    Flavours II}, eds. A.J. Buras and M. Lindner, World Scientific
  Publishing Co., Singapore 1998 ({\sl hep-ph}/9703442).
  
\bibitem{BUGAGOJASI2} A.J. Buras, P. Gambino, M. Gorbahn, S. J\"ager
  and L. Silvestrini, {\sl Nucl. Phys.} {\bf B592} (2001), 55.
  
\bibitem{BUCHROSL} A.J. Buras, P.H. Chankowski, J. Rosiek and {\L}.
  S{\l}awianowska, in preparation.
  
\bibitem{BUJAWE} A.J. Buras, M. Jamin, and P.H. Weisz, {\sl Nucl.
    Phys.}  {\bf B347} (1990), 491.
  
\bibitem{URKRJESO} J. Urban, F. Krauss, U. Jentschura and G. Soff,
  {\sl Nucl. Phys.} {\bf B523} (1998) 40.
  
\bibitem{CET0} M. Ciuchini, E. Franco, V. Lubicz, G. Martinelli, I.
  Scimemi and L. Silvestrini, {\sl Nucl. Phys.} {\bf B523} (1998) 501.
  
\bibitem{BUMIUR} A.J. Buras, M. Misiak and J. Urban, {\sl Nucl. Phys.}
  {\bf B586} (2000), 397.
  
\bibitem{CET} M. Ciuchini, et al., {\sl JHEP} {\bf 9810} (1998), 008.
  
\bibitem{BUJAUR} A.J. Buras, S. J\"ager and J. Urban, {\sl Nucl.
    Phys.} {\bf B605} (2001), 600.
  
\bibitem{WO} L. Wolfenstein, {\sl Phys. Rev. Lett.} {\bf 51} (1983),
  1945.
  
\bibitem{HENI} S. Herrlich and U. Nierste, {\sl Nucl. Phys.} {\bf
    B419} (1994), 292; {\sl Phys. Rev.} {\bf D52} (1995) 6505; {\sl
    Nucl. Phys.} {\bf B476} (1996), 27.
  
\bibitem{FL} R. Fleischer, {\sl Phys. Lett.} {\bf B459} (1999), 306.
  
\bibitem{BUFL2} A.J. Buras and R. Fleischer, {\sl Eur. Phys. J.}  {\bf
    C11} (1999) 93; {\sl Eur. Phys. J.} {\bf C16} (2000) 97; {\sl
    hep-ph}/0008298.
  
\bibitem{HEHOUYA} X.-G. He, W.-S. Hou and K-Ch. Yang, {\sl
    hep-ph}/9902256; W.-S. Hou and K.-Ch. Yang, {\sl Phys. Rev.} {\bf
    D61} (2000) 073014; W.-S. Hou, J.G. Smith and F. W\"urthwein, {\sl
    hep-ex}/9910014.
  
\bibitem{BEBENESA} M. Beneke, G. Buchalla, M. Neubert and C.T.
  Sachrajda, {\sl hep-ph}/0007256; {\sl hep-ph}/0104110;
  
\bibitem{DUYAZH} D. Du, D. Yang and G. Zhu, {\sl hep-ph}/0005006; T.
  Muta, A. Sugamoto, M.Z. Yang and Y.D. Yang, {\sl hep-ph}/0006022.
  
\bibitem{CIFRMAPISI} M. Ciuchini, E. Franco, G. Martinelli, M. Pierini
  and L. Silvestrini, {\sl hep-ph}/0104126.
  
\bibitem{latt_lit} C. R. Allton et al., {\sl Phys. Lett.}  {\bf B453}
  (1999), 30.
  
\bibitem{BAMAZH} J.A.~Bagger, K.T.~Matchev and R.J.~Zhang, {\sl Phys.
    Lett.}  {\bf B412} (1997), 77.
  
\bibitem{FLLISA} J. Flynn and C.-J.D. Lin, talk given at UK
  Phenomenology Workshop on Heavy Flavor and CP Violation, Durham,
  England, Sepember 2000, {\sl hep-ph}/0012154; C.T. Sachrajda, {\sl
    hep-lat}/0101003.
  
\bibitem{ROS} J. Rosiek, {\sl Phys. Rev.} {\bf D41} (1990), 3464, {\sl
    Erratum} {\sl hep-ph}/9511250.
  
\bibitem{GAMI} P. Gambino and M. Misiak, preprint CERN-TH/2000-089
  ({\sl hep-ph}/0104034);
  
\bibitem{CLEOMOR} F. Blanc (the CLEO Collaboration), talk at the
  XXXVI$^{th}$ Rencontres de Moriond, March 2001, Les Arcs 1800,
  France.
  
\bibitem{GAMOR} P. Gambino, talk at the XXXVI$^{th}$ Rencontres de
  Moriond, March 2001, Les Arcs 1800, France.
  
\bibitem{GAGI} E. Gabrielli and G.-F. Giudice, {\sl Nucl. Phys.}  {\bf
    B433} (1995), 3.
  
\bibitem{BR} N.H. Brown et al., {\sl Phys. Rev. Lett.} {\bf 86}
  (2001), 2227.
  
\bibitem{DEGAGI} G. Degrassi, P. Gambino and G.-F. Giudice, {\sl JHEP}
  {\bf 0012} (2000), 009.
  
\bibitem{CAGANIWA} M. Carena, D. Garcia, U. Nierste and C.E.M. Wagner,
  {\sl Phys. Lett.} {\bf B499} (2001), 141.
  
\bibitem{CHPO} P.H. Chankowski and S. Pokorski in {\sl Perspectives on
    supersymmetry}, G.L. Kane ed., World Scientific Publishing Co.,
  Singapore 1998 ({\sl hep-ph}/9707497); M. Carena, U. Nierste and C.E.M.
  Wagner, {\sl Nucl. Phys.} {\bf B577} (2000), 88.
  
\bibitem{HARASA} L.J. Hall, R. Rattazzi and U. Sarid, {\sl Phys. Rev.}
  {\bf D50} (1994) 7048.
  
\bibitem{HAPOTO} C. Hamzaoui, M. Pospelov and M. Toharia, {\sl Phys.
    Rev.}  {\bf D59} (1999), 095005.
  
\bibitem{BAKO} K.S. Babu and C. Kolda, {\sl Phys. Rev. Lett.}  {\bf
    84} (2000), 228.
  
\bibitem{CHSL} P.H. Chankowski and {\L}. S{\l}awianowska {\sl Phys.  Rev.}
  {\bf D63} (2001), 054012.
  
\bibitem{BOEWKRUR} C. Bobeth, T. Ewerth, F. Kr\"uger and J. Urban,
  preprint TUM-411/01, {\sl hep-ph}/0104284.
  
\bibitem{HULIYAZH} C.-S. Huang, W. Liao, Q.-S. Yan and S.-H. Zhu, {\sl
    Phys. Rev.} {\bf D63} (2001), 114021, Erratum {\sl ibid.} {\bf
    D64} (2001), 059902.

\end{thebibliography}
\end{document}